\documentclass[AMA,STIX1COL]{WileyNJD-v2}

\articletype{Article Type}%

\received{Date Month Year}
\revised{Date Month Year}
\accepted{Date Month Year}

\copyyear{2023}
\startpage{1}

\usepackage{ifpdf}
\usepackage{color, colortbl}
\usepackage{tikz}
\usepackage{array}
\usepackage{multirow}
\usepackage{listings}
\usepackage{xcolor}
\usepackage{amsmath}
\usepackage{graphicx}
\usepackage{subcaption}
\usepackage{url}
\usepackage{booktabs}
\usepackage{pifont}
\usepackage{adjustbox,siunitx}
\usepackage{booktabs}
\usepackage{tabularx}

\lstdefinelanguage{json}{
    basicstyle=\normalfont\ttfamily,
    stringstyle=\color{darkgreen},
    literate=
     *{0}{{{\color{numb}0}}}{1}
      {:}{{{\color{punct}{:}}}}{1},
    morestring=[b]"
}
\colorlet{punct}{red!60!black}
\definecolor{background}{HTML}{EEEEEE}
\definecolor{darkgreen}{HTML}{006400}
\definecolor{delim}{RGB}{20,105,176}
\colorlet{numb}{magenta!60!black}

\begin{document}

\setcounter{page}{1}

\title{Cloud-native and Distributed Systems for Efficient and Scalable Large Language Models -- A Research Agenda}

\author[1]{Minxian Xu}

\author[1]{Jingfeng Wu}

\author[1]{Shengye Song}

\author[2]{Satish Narayana Srirama}

\author[3]{Bahman Javadi}

\author[4]{Rajiv Ranjan}

\author[4]{Devki Nandan Jha}

\author[5]{Sa Wang}

\author[6]{Wenhong Tian}

\author[7]{Huanle Xu}

\author[7]{Li Li}

\author[7]{Zizhao Mo}

\author[8]{Shuo Ren}

\author[9]{Thomas Kunz}

\author[10]{Petar Kochovski}

\author[10]{Vlado Stankovski}

\author[1]{Kejiang Ye}

\author[7]{Chengzhong Xu}

\author[12]{Rajkumar Buyya}

\authormark{Xu \textsc{et al.}}
\titlemark{Cloud-native and Distributed Systems for LLMs}

\address[1]{\orgdiv{Shenzhen Institutes of Advanced Technology}, \orgname{Chinese Academy of Sciences}, \orgaddress{\state{Shenzhen}, \country{China}}}

\address[2]{\orgdiv{University of Hyderabad}, \orgaddress{\state{Hyderabad}, \country{India}}}

\address[3]{\orgdiv{Western Sydney University}, \orgaddress{\state{Sydney}, \country{Australia}}}

\address[4]{\orgdiv{Newcastle University}, \orgaddress{\state{Newcastle}, \country{United Kingdom}}}

\address[5]{\orgdiv{Institute of Computing Technology, Chinese Academy of Sciences}, \orgaddress{\state{Beijing}, \country{China}}}

\address[6]{\orgdiv{University of Electronic Science and Technology of China}, \orgaddress{\state{Chengdu}, \country{China}}}

\address[7]{\orgdiv{University of Macau}, \orgaddress{\state{Macau SAR}, \country{China}}}

\address[8]{\orgdiv{Institute of Automation, Chinese Academy of Sciences}, \orgaddress{\state{Beijing}, \country{China}}}

\address[9]{\orgdiv{Carleton University}, \orgaddress{\state{Ottawa}, \country{Canada}}}

\address[10]{\orgdiv{University of Ljubljana}, \orgaddress{\state{Ljubljana}, \country{Slovenia}}}


\address[12]{\orgdiv{University of Melbourne}, \orgaddress{\state{Melbourne}, \country{Australia}}}


\corres{Minxian Xu, Shenzhen Institutes of Advanced Technology, Chinese Academy of Sciences, China \\ \email{mx.xu@siat.ac.cn}}

\abstract[Abstract]{
The rapid rise of Large Language Models (LLMs) has revolutionized various artificial intelligence (AI) applications, from natural language processing to code generation. However, the computational demands of these models, particularly in training and inference, present significant challenges. Traditional systems are often unable to meet these requirements, necessitating the integration of cloud-native and distributed architectures. This paper explores the role of cloud platforms and distributed systems in supporting the scalability, efficiency, and optimization of LLMs. We discuss the complexities of LLM deployment, including data management, resource optimization, and the need for microservices, autoscaling, and hybrid cloud-edge solutions. Additionally, we examine emerging research trends, such as serverless inference, quantum computing, and federated learning, and their potential to drive the next phase of LLM innovation. The paper concludes with a roadmap for future developments, emphasizing the need for continued research, standardization, and cross-sector collaboration to sustain the growth of LLMs in both research and enterprise applications.
}


\maketitle

\section{Introduction}\label{sec:introduction}

LLMs have rapidly emerged as a transformative force in AI, reshaping how machines perceive, generate, and reason over human language. In just a few years, LLMs have driven breakthroughs across a wide spectrum of applications, including conversational agents, code generation, scientific discovery, content creation, and enterprise automation\cite{distserve}. Their unprecedented capability to generalize across tasks with minimal task-specific training has fundamentally altered both research paradigms and industrial AI development, positioning LLMs as a foundational component of modern digital infrastructure rather than isolated application-level models.
The rise of LLMs also marks a significant departure from traditional machine learning (ML) and deep learning (DL) models. Conventional ML systems typically rely on relatively compact models designed for narrowly scoped tasks, with predictable training and inference workloads. Even earlier deep learning models, such as CNNs for vision or RNNs for sequence processing, were often deployed in relatively static environments with well-understood resource requirements. In contrast, LLMs are characterized by massive parameter scales, data-hungry training pipelines, and highly dynamic inference patterns. Their training often spans thousands of GPUs or specialized accelerators over extended periods, while inference workloads exhibit burstiness, variable sequence lengths, multi-tenancy, and increasingly interactive or agentic behaviors. These differences fundamentally challenge existing assumptions in system design, scheduling, and resource management.

While LLMs promise unprecedented intelligence and productivity gains, they also introduce severe challenges for AI infrastructure (like cloud and distributed systems that host LLMs). The computational, memory, storage, and networking demands of LLM training and inference significantly stress current systems. Inefficient resource utilization can lead to escalating operational costs, energy consumption, and carbon footprints, raising sustainability concerns. Moreover, the growing diversity of deployment scenarios, ranging from hyperscale cloud data centers to edge devices and hybrid environments, complicates resource provisioning and orchestration. Ensuring predictable latency, high throughput, fairness among tenants, and fault tolerance under such heterogeneous and dynamic workloads remains an open problem. As LLM-based services increasingly become mission-critical, resource management is no longer merely an optimization problem but a fundamental requirement for reliability, scalability, and economic viability.

Cloud-native systems and distributed systems offer a natural and powerful foundation for addressing these challenges. Technologies such as microservices, containerization, orchestration platforms, autoscaling mechanisms, and service meshes enable flexible, scalable, and fault-tolerant deployment of complex applications\cite{container}. When combined with distributed system principles, such as parallelism, data locality, consistency management, and decentralized control, cloud-native infrastructures can provide the elasticity and efficiency required by large-scale LLM workloads. However, LLMs also expose limitations in existing cloud-native abstractions, calling for new system designs that are LLM-aware in terms of computation, communication, and energy efficiency. This interplay between LLM workloads and system infrastructure creates rich opportunities for innovation at the intersection of AI and systems research.

In this context, research agenda and vision papers play a critical role in guiding and organizing community-wide efforts. Rather than proposing a single system or algorithm, such works aim to synthesize emerging trends, identify fundamental challenges, and articulate open research questions that can shape future exploration. Given the fast-evolving nature of LLM technologies and their deep entanglement with system infrastructure, there is an urgent need for a coherent research agenda that bridges cloud-native systems, distributed systems, and large-scale AI. This paper seeks to contribute to this goal by outlining key challenges, design principles, and research directions for efficient resource management of LLMs, with the aim of fostering cross-disciplinary collaboration and accelerating progress in this rapidly growing field.
At the time of archiving this work, no vision paper specifically focusing on cloud-native and distributed system support of LLMs was available. Given the rapid evolution of this field, several related papers have since emerged. However, none of them present a comprehensive research agenda that aligns with the scope and intent of our work.  

\color{blue}Existing studies have extensively investigated different aspects of large language
models, including model architectures, distributed training systems, inference
optimization, and efficient serving frameworks. However, most existing surveys focus
on individual components of the LLM lifecycle rather than the holistic evolution of
cloud-native and distributed infrastructures required to support large-scale LLM
services. Table~\ref{tab:related_work_comparison} summarizes the differences between
our work and representative surveys and vision papers. Unlike prior studies, this
work adopts an infrastructure-centric perspective and investigates the interactions
among LLM workloads, cloud-native software stacks, resource orchestration mechanisms,
and future AI-native computing paradigms.

\begin{table*}[t]
	\centering
	\caption{\color{blue}Comparison between this work and representative
		existing surveys on LLM systems and AI infrastructure.}
	\color{blue}	\noindent\footnotesize
		\textit{Note:} $\times$ indicates that the topic is largely outside the scope of
		the work; ``Limited'' indicates only incidental discussion; ``Partial'' indicates
		that the topic is considered but is not a primary organizing dimension;
		\checkmark indicates substantial coverage; and \checkmark\checkmark indicates
		that the topic constitutes a central focus of the work.
	\label{tab:related_work_comparison}
     {\color{blue}
	\resizebox{\textwidth}{!}{
		\begin{tabular}{lcccccc}
			\toprule
			\textbf{Work} &
			\textbf{Main Focus} &
			\textbf{Training Infrastructure} &
			\textbf{LLM Serving} &
			\textbf{Cloud-native Systems} &
			\textbf{Resource Management} &
			\textbf{Research Agenda} \\
			\midrule
			
			Zhao et al.~\cite{zhao2026survey}
			& General LLM techniques
			& Partial
			& Partial
			& $\times$
			& Limited
			& Partial
			\\
			
			Duan et al.~\cite{duan2026efficient}
			& Distributed LLM training infrastructure
			& \checkmark\checkmark
			& $\times$
			& Limited
			& \checkmark
			& Partial
			\\
			
			Li et al.~\cite{li2024llmserving}
			& LLM inference and serving systems
			& $\times$
			& \checkmark\checkmark
			& Partial
			& \checkmark\checkmark
			& Partial
			\\
			
			Wan et al.~\cite{wan2024efficient}
			& Efficient LLMs
			& \checkmark
			& \checkmark
			& Limited
			& Partial
			& Partial
			\\
			
			Zheng et al.~\cite{zheng2025edge}
			& Edge LLM design and execution
			& Limited
			& \checkmark
			& $\times$
			& \checkmark
			& Partial
			\\
			
			\textbf{This work}
			& \textbf{Cloud-native and distributed LLM infrastructures}
			& \textbf{\checkmark}
			& \textbf{\checkmark\checkmark}
			& \textbf{\checkmark\checkmark}
			& \textbf{\checkmark\checkmark}
			& \textbf{\checkmark\checkmark}
			\\
			
			\bottomrule
		\end{tabular}
	}}
\end{table*}

Moreover, as shown in Table~\ref{tab:related_work_comparison}, existing surveys provide
valuable insights into specific aspects of LLM development, such as distributed
training or inference optimization. In contrast, this work aims to establish a
broader research agenda covering the complete cloud-native infrastructure stack for
LLM services, including heterogeneous hardware management, distributed execution,
elastic resource orchestration, multi-tenant service management, security, and
future intelligent computing paradigms.\color{black}

To provide a coherent research roadmap, this paper organizes the discussion from challenges to solutions and future opportunities. We first analyze why existing cloud-native and distributed infrastructures cannot efficiently support emerging LLM workloads. We then review representative system-level solutions addressing these challenges, including resource orchestration, distributed execution, and intelligent management mechanisms. Finally, we discuss emerging paradigms and future research directions toward next-generation AI-native infrastructures.

The rest of the paper is structured as follows: we discuss the relevant LLM background and the motivation that cloud-native and distributed system to support LLM  in Section \ref{ch:2}, and present the challenges to support LLMs from the system perspective in Section \ref{ch:3}. Section \ref{ch:4} highlights how cloud-native and distributed systems can achieve scalable and efficient resource management for LLMs. Moreover, Section \ref{ch:5} discusses emerging trends in this area and Section \ref{ch:6} outlines the future research agenda. Section \ref{ch:7} summarizes the key findings in this work, and Section \ref{ch:8} concludes the paper.
\color{blue}\section{Background}\label{ch:2}\color{black}

The application of AI and ML within large-scale computing systems has a long research history, particularly in areas such as distributed scheduling, workload prediction, and resource optimization. Traditional DL systems have been widely deployed across domains such as computer vision, speech recognition, and recommendation systems, where models are typically task-specific, relatively stable after deployment, and operate under well-defined input-output patterns. These workloads are often optimized for high-throughput batch processing or narrowly scoped inference, allowing existing distributed infrastructures to scale efficiently using established parallelization and scheduling techniques. In contrast, LLMs introduce a fundamentally different paradigm. They are general-purpose and highly expressive, capable of handling diverse tasks through a unified interface. Their autoregressive, token-level generation process leads to iterative execution patterns, dynamic control flow, and strong dependence on intermediate states such as key-value caches, which significantly alters system behavior.

Beyond execution characteristics, LLMs also differ from traditional DL models in terms of scale, adaptability, and system implications. Modern LLMs contain billions or even trillions of parameters, placing unprecedented demands on memory capacity, communication efficiency, and distributed coordination. At the same time, emerging techniques such as in-context learning, prompt-based interaction, and parameter-efficient fine-tuning enable continuous adaptation without full retraining, blurring the boundary between training and inference and introducing new challenges in multi-tenancy, model lifecycle management, and resource isolation. The integration of LLMs into production systems therefore represents a rapidly evolving shift. While earlier neural language models and transformer architectures established the technical foundation, the widespread adoption of services such as ChatGPT, DeepSeek, and Codex has accelerated their transition into large-scale, user-facing systems. This evolution calls for a rethinking of distributed system design, motivating the development of LLM-aware, cloud-native infrastructures capable of supporting their unique computational and operational requirements.

\begin{figure}
    \centering
    \includegraphics[width=0.8\linewidth]{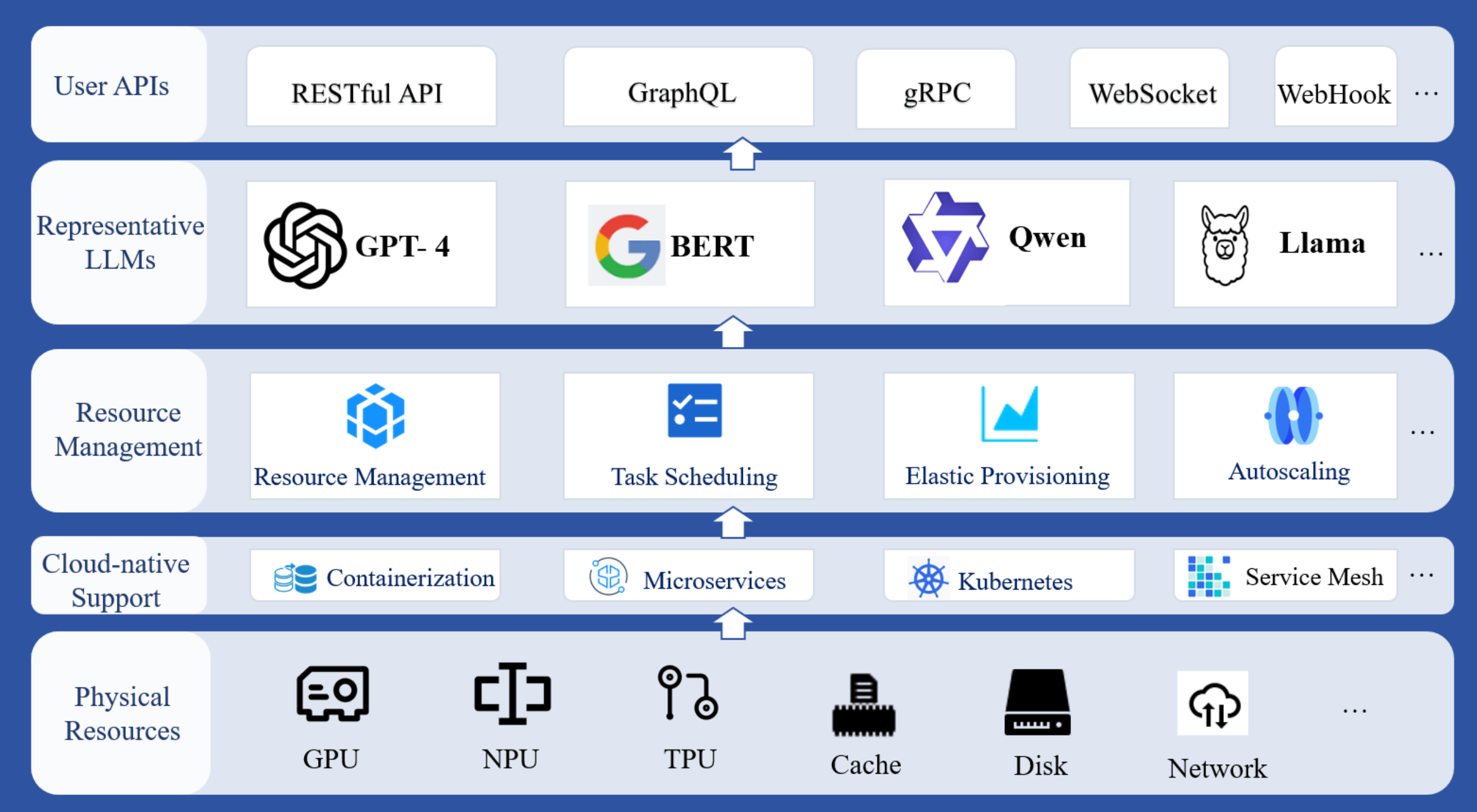}
    \caption{Cloud-native system support for LLMs}
    \label{fig:systemarch}
\end{figure}

Fig. \ref{fig:systemarch} presents a layered architecture for a cloud-native system supporting LLMs, spanning from underlying hardware to user-facing interfaces. At the foundation, heterogeneous physical resources including GPUs, NPUs, TPUs, storage, and networking provide the essential computational substrate. On top of this, the cloud-native support layer, built on technologies such as containerization, microservices, Kubernetes, and service mesh, enables scalable and modular deployment across distributed environments. The resource management layer introduces dynamic system intelligence through task scheduling, elastic provisioning, and autoscaling to efficiently handle the highly variable and resource-intensive nature of LLM workloads. Above this, representative LLMs such as GPT-4, BERT, Qwen, and LLaMA are deployed and optimized within the system stack, while the top layer exposes standardized APIs including RESTful, GraphQL, gRPC, WebSocket, and WebHook to facilitate seamless integration with downstream applications. Together, these layers illustrate how cloud-native and distributed system principles jointly enable scalable, flexible, and efficient LLM services.

As LLM capabilities have improved, research attention has expanded beyond model accuracy and architecture design to encompass system-level challenges, including scalability, efficiency, cost control, and reliability. A growing body of work, now widely being disseminated through platforms such as open-source ecosystems, has begun to investigate how cloud-native and distributed infrastructures can effectively support LLM training and inference. To ground the subsequent research agenda, this section introduces relevant terminologies (Section \ref{sec 2.1}), outlines the historical evolution and fundamental properties of LLMs (Section \ref{sec 2.2}), and reviews existing system support approaches for LLM workloads (Section \ref{sec 2.3}).

\subsection{Terminologies} \label{sec 2.1}

LLMs generally refer to deep neural networks trained on massive corpora of textual (and increasingly multimodal with images, videos and audio data) data to perform language understanding and generation tasks. Most contemporary LLMs are based on the transformer architecture, which processes input in the form of tokens and relies on self-attention mechanisms to model long-range dependencies of already appeared tokens and enable scalable parallel training\cite{transformer}. After attention, the transformer layer includes a multilayer perceptron block, which further refines the token embeddings to include complex feature transformations. The LLM model includes several layers of such self-attention and perceptron blocks, e.g. up to 96 layers in the case of GPT-3 model variants. 

Key concepts associated with LLMs include pretraining, where models learn general linguistic representations from large datasets, and fine-tuning or instruction tuning, which adapts pretrained models to downstream tasks or interaction paradigms such as text generation, summarization, question answering and providing comprehensive assessment in the case of multimodal data.

From a systems perspective, several additional terminologies are essential. Inference serving denotes the process of deploying trained LLMs to respond to user requests, often under strict latency and throughput constraints\cite{Zheng2025SmartData}. Batching aggregates multiple inference requests to improve hardware utilization, while model parallelism and data parallelism distribute computation across multiple devices or nodes. Elasticity and autoscaling refer to a system’s ability to dynamically adjust resource allocation in response to workload fluctuations. In cloud-native environments, concepts such as containers, microservices, and orchestration platforms (e.g., Kubernetes) play a central role in managing LLM workloads.

Furthermore, emerging terms such as LLM-as-a-Service, serverless inference, and hybrid cloud–edge deployment reflect evolving deployment paradigms. These terminologies highlight a shift from static, monolithic model hosting toward flexible, on-demand, on-device and distributed execution models, which place new demands on resource management and system coordination.

\subsection{Historical evolution and fundamentals of LLMs}
\label{sec 2.2}
The development of LLMs can be traced back to early statistical language models and neural network–based approaches, including recurrent neural networks and sequence-to-sequence models. The introduction of the transformer architecture marked a turning point by enabling efficient parallelization and scalable training on large datasets~\cite{transformer}. Subsequent increases in model size, training data, and computational resources gave rise to models with billions or even trillions of parameters, demonstrating strong emergent capabilities such as in-context learning, where the LLMs learn to perform new tasks by including examples directly in the prompt, without updating model weights, and zero-shot generalization where the model correctly performs AI tasks on data it has never encountered during training.

A defining characteristic of modern LLMs is their tight coupling between model performance and computational scale. Empirical scaling laws have shown that improvements in accuracy and generalization often require proportional increases in training data and compute resources~\cite{scaling_laws}. As a result, LLM development has become deeply intertwined with large-scale distributed training infrastructures and specialized hardware accelerators. On the inference side, the interactive and often unpredictable nature of user queries introduces additional complexity, as request sizes, sequence lengths, and response times can vary widely.

These fundamentals distinguish LLMs from earlier ML and DL models not only in terms of capability but also in their system-level behavior. The heavy reliance on memory bandwidth, high-speed interconnects, and synchronization across devices makes LLM workloads particularly sensitive to system design choices. Understanding these characteristics is essential for designing efficient cloud-native and distributed support mechanisms.

Furthermore, for inference on resource-constrained devices across the edge-fog-cloud continuum, optimizations such as parameter-weight quantization, pruning, low-rank adaptation (LoRA) etc. are developed. Parameter weight quantization is a model compression technique that reduces the precision of the LLM weights from 32-bit floating point (FP32) to lower-precision formats such as INT8, INT4, to decrease memory usage and accelerate inference, often with minimal loss in accuracy. Similarly, pruning optimizes LLMs by identifying and removing redundant or less important parameters. LoRA performs task-specific adaptation by adding a trainable low-rank update matrix, represented by the product of two smaller matrices, to the model’s original frozen weights.

\subsection{System support for LLMs} \label{sec 2.3}
To meet the demanding requirements of LLM training and inference, researchers and practitioners have increasingly turned to cloud-native and distributed system solutions. Distributed training frameworks leverage parallelism across GPUs and nodes to reduce training time, while checkpointing and fault-tolerance mechanisms ensure robustness at scale. For inference, systems must balance latency, throughput, and cost, often under multi-tenant conditions with highly dynamic workloads. Furthermore, the possibility of inference across the edge-fog-cloud continuum makes the system support for LLMs a lot more complex.

Cloud-native abstractions such as containerized services, microservice architectures, and orchestration frameworks provide a flexible foundation for deploying LLM services. Autoscaling policies can adjust resource allocation based on demand, while service meshes and load balancers manage request routing and isolation. At the same time, new challenges arise in efficiently managing large model states, minimizing communication overhead, and coordinating resources across heterogeneous environments, including edge and hybrid cloud platforms.

Despite significant progress, existing system support mechanisms are often adapted from general-purpose cloud workloads and are not explicitly designed with LLM characteristics in mind. This mismatch motivates further research into LLM-aware system designs that integrate model behavior, workload dynamics, and resource management policies. Addressing these gaps is central to enabling sustainable, scalable, and efficient LLM deployments, and it forms the foundation for the research agenda outlined in the subsequent sections of this paper.

\color{blue}
\subsection{Methodology of constructing the research agenda} \label{sec 2.4}
To develop a comprehensive research agenda for cloud-native and distributed systems
supporting LLMs, we conduct a structured analysis of representative research
advances and practical developments across multiple related areas, including
large-scale distributed training, LLM inference and serving, cloud-native
orchestration, resource management, heterogeneous computing, and edge-cloud
intelligence. Rather than performing a purely literature-counting survey, we
identify key research challenges based on three criteria: (1) whether the challenge
is introduced by the unique characteristics of LLM workloads, such as large model
scale, dynamic execution patterns, and intensive resource requirements; (2) whether
addressing the challenge requires fundamental advances in cloud-native and
distributed system technologies; and (3) whether the research direction has
significant impact on the scalability, efficiency, reliability, and sustainability
of future AI services. Following these criteria, we organize the research agenda
from fundamental system challenges, to existing solutions, and finally to emerging
opportunities for next-generation AI-native infrastructures.\color{black}

 
\section{Challenges in Supporting LLMs from Systems}
\label{ch:3}

\begin{figure}
	\centering
	\includegraphics[width=0.65\linewidth]{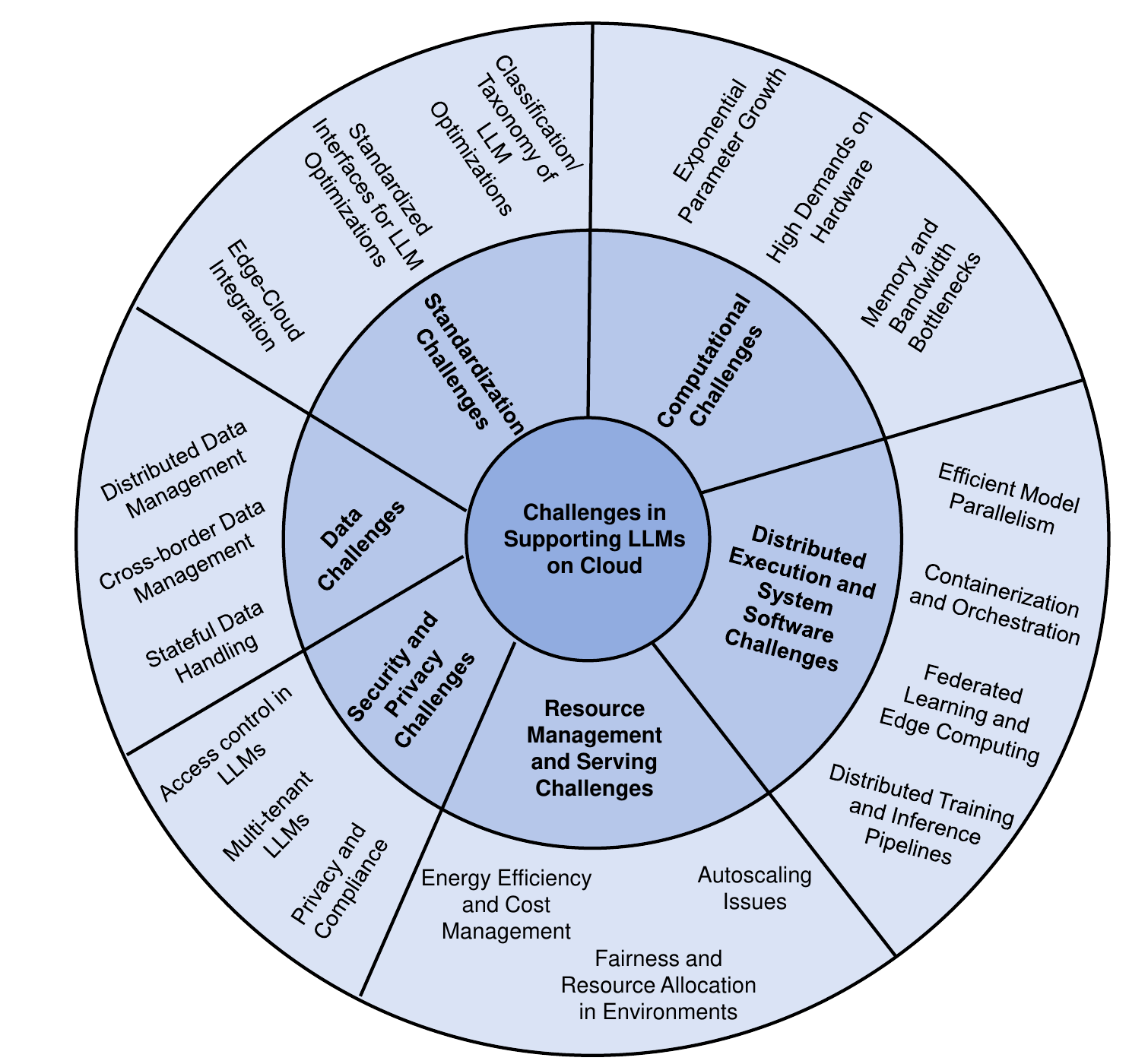}
	\caption{\color{blue}Challenges in Supporting LLMs from Systems}
	\label{fig:challenges}
\end{figure}

This section focuses on the various challenges faced when supporting LLMs from a system perspective, covering computational, system, operational, and data challenges, as well as the need for standardization. Despite rapid advances in model architectures and hardware accelerators, efficiently supporting LLM workloads at scale remains a major challenge for system designers. LLMs fundamentally alter the assumptions underlying traditional AI systems, introducing new computational patterns, software complexity, operational constraints, and societal concerns. This section outlines the key challenges in system support for LLMs, spanning computation, software infrastructure, operations, privacy, data, and standardization, and highlights open research problems in each dimension as shown in Fig. \ref{fig:challenges}.

\color{blue}To distinguish the research agenda developed in this paper from research questions
addressed by an individual technical study, we use the term \textit{Open Research
	Questions} throughout this section. These questions are not intended to be answered
by the present paper; rather, they are derived from the challenges, limitations, and
gaps identified in the existing literature and are intended to highlight promising
opportunities for future research. For each challenge dimension, we first examine
the underlying characteristics and current state of the field, and then formulate
open research questions that connect the identified limitations to potential future
investigations. In this way, the questions serve as a bridge from the current state
of LLM systems to the broader research agenda developed in the subsequent sections.\color{black}

\subsection{Computational Challenges}
\label{sec:computational-challenges}

\subsubsection{Historical Context}

Early ML systems were predominantly CPU-bound, with workloads characterized by relatively small datasets and models that could be trained and deployed within centralized environments. Resource management concerns were minimal, and scheduling strategies were largely static due to predictable computational requirements. The emergence of DL marked a significant shift toward accelerator-based computation, particularly with GPUs, which enabled efficient execution of dense linear algebra operations~\cite{lecun2015deep}. Frameworks such as TensorFlow~\footnote{\href{https://www.google.com/url?q=https://www.tensorflow.org/&sa=D&source=docs&ust=1774321803274743&usg=AOvVaw2bBtlzGy0dhw3WZWUMcpd3}{https://www.tensorflow.org/}} and PyTorch~\footnote{\href{https://www.google.com/url?q=https://pytorch.org/&sa=D&source=docs&ust=1774321803274909&usg=AOvVaw3szUUnW0FtEuZjL9jlHDP5}{https://pytorch.org/}} further abstracted hardware complexities, allowing developers to scale training across multiple devices. Despite these advances, deep neural networks prior to LLMs remained manageable in scale and exhibited relatively stable computational patterns.

The introduction of transformer architectures and subsequent scaling laws fundamentally altered this landscape. Models such as Generative Pre-trained Transformer (GPT), Pathways Language Model (PaLM), and Large Language Model Meta AI (LLaMA) introduced parameter counts ranging from billions to trillions, along with long-sequence processing requirements and autoregressive decoding behaviors that are inherently sequential at inference time~\cite{isaev2023scaling}. This shift has led to a transition from coarse-grained parallelism (e.g., data parallelism) to fine-grained, multi-dimensional parallelism, including tensor, pipeline, sequence, and expert parallelism. Moreover, LLM workloads exhibit high variability across requests (e.g., input/output length, decoding strategy), making static allocation strategies inefficient. As a result, computation is no longer merely a function of model size but also of runtime dynamics, system topology, and workload heterogeneity.

\subsubsection{Open Research Questions}
\color{blue}
Based on the computational challenges and limitations discussed above, we identify
the following open research questions for future LLM system research:

\begin{itemize}
    \item RQ1: How can cloud-native systems achieve adaptive computation optimization for heterogeneous LLM workloads?
    \item RQ2: How can distributed LLM execution efficiently coordinate computation, memory, and communication across heterogeneous accelerators?
    \item RQ3: How can AI-driven optimization techniques enable autonomous resource management for LLM computation?
\end{itemize}\color{black}

\subsubsection{State of the Art}

Recent advances in LLM computation spans advances in distributed training, inference optimization, and hardware-software co-design. A central theme across these developments is the need to balance computational efficiency with scalability under increasingly complex workloads. Distributed training of LLMs relies heavily on hybrid parallelism techniques. Data parallelism replicates model parameters across devices and aggregates gradients, but incurs communication overheads that grow with scale. Tensor parallelism partitions individual layers across multiple devices, reducing memory pressure while increasing communication frequency. Pipeline parallelism divides models into stages executed across devices, improving utilization but introducing scheduling challenges such as pipeline bubbles. Systems such as Megatron-LM~\cite{megatronlm} and SpeedLoader~\cite{speedloader} demonstrate how these techniques can be combined to train multi-billion parameter models efficiently, though their performance remains highly sensitive to system topology and interconnect bandwidth.

On the inference side, optimization techniques focus on reducing latency and improving throughput under real-time constraints. Dynamic batching aggregates requests to increase hardware utilization, though it introduces trade-offs between latency and efficiency. Mixed-precision execution leverages lower numerical precision to reduce computational cost, while quantization and sparsity techniques further compress models to enable efficient deployment~\cite{aminabadi2022deepspeed}. More recently, speculative decoding has emerged as a promising approach to accelerate autoregressive generation by leveraging smaller draft models, reducing end-to-end latency when predictions are accurate. Additionally, optimized attention mechanisms such as FlashAttention~\cite{flashattention} significantly improve memory efficiency and computational speed by reducing memory movement during attention computation. Another important direction is hardware-software co-design, where system optimizations are tightly integrated with hardware capabilities. Optimization frameworks such as Ayo~\cite{ayo} enables task primitives as the basic units and represents each query's workflow as a primitive-level dataflow graph. Similarly, domain-specific accelerators such as TPUs~\cite{tpu} are designed to optimize tensor operations, delivering significant performance gains for large-scale workloads. Systems like Alpa~\cite{alpa} further automate parallelization strategies, abstracting away low-level system complexities while optimizing execution across distributed environments.

\subsubsection{Challenges and Limitations}

Despite these advances, current state-of-the-art solutions often assume relatively static workloads and homogeneous hardware environments. Many optimizations are tailored to specific models or deployment scenarios, limiting their generalizability. As a result, applying these techniques to dynamic, multi-tenant cloud LLM environments remains a significant challenge.

A fundamental challenge in LLM computation lies in handling the high degree of workload variability observed in real-world deployments. Unlike traditional batch-processing systems, LLM inference must accommodate diverse request patterns, including variations in input length, output generation, and latency requirements. Existing optimization techniques, such as static batching or fixed parallelism strategies, are ill-suited to such dynamic environments, often resulting in suboptimal resource utilization or degraded quality of service.

Communication overhead represents another critical bottleneck in distributed LLM systems. As models are partitioned across multiple devices, the need for frequent synchronization and data exchange can dominate overall execution time. This is particularly evident in tensor and pipeline parallelism, where inter-device communication is tightly coupled with computation. The performance of such systems is therefore highly dependent on interconnect technologies, and inefficiencies in communication can negate the benefits of parallelism~\cite{megatronlm}.

Memory constraints further add to the computational challenges. The size of modern LLMs often exceeds the memory capacity of individual accelerators, necessitating model sharding, activation checkpointing, or offloading to host memory. During inference, the growth of key-value caches for autoregressive decoding imposes additional memory pressure, limiting scalability and increasing latency. Although techniques such as memory-efficient attention mitigate some of these issues, memory bandwidth remains a dominant constraint in many scenarios~\cite{vllm}.

Heterogeneity in hardware resources introduces additional complexity. Cloud environments increasingly consist of diverse accelerators with varying performance characteristics, memory capacities, and interconnect capabilities. Efficiently utilizing such heterogeneous resources requires fine-grained scheduling and hardware-aware optimization, yet most existing frameworks provide limited support for these capabilities. This leads to resource fragmentation and underutilization in practice. Energy efficiency is an emerging concern as LLM workloads scale. Training and inference at scale consume substantial amounts of energy, raising both economic and environmental challenges. Current systems often prioritize performance over energy efficiency, leading to overprovisioning and inefficient resource usage. The lack of energy-aware scheduling mechanisms further limits the ability to optimize for sustainability.

Finally, there is a lack of generalizable system abstractions for LLM computation. Many existing solutions are tightly coupled to specific models, hardware platforms, or frameworks, making them difficult to adapt to new environments. This fragmentation hinders the development of portable and reusable optimization strategies, increasing system complexity and limiting broader adoption.

\subsubsection{Future Prospects}

Cloud-native and distributed systems are expected to play a central role in enabling efficient and scalable LLM computation. Rather than acting as passive execution substrates, future systems are likely to evolve into intelligent infrastructures that actively manage and optimize computational resources across the model lifecycle.

One promising direction is the development of fine-grained scheduling mechanisms that operate at the level of individual tokens or operators. Such approaches would enable more precise allocation of computational resources, improving utilization and reducing latency in dynamic environments. Coupled with advances in distributed execution frameworks, this could enable seamless computation across cloud and edge resources.

Adaptive and self-optimizing systems represent another key avenue for future research. By incorporating learning-based controllers, system runtimes can dynamically adjust parameters such as batching strategies, precision levels, and parallelism configurations in response to changing workload conditions. This would enable more efficient and robust operation in multi-tenant environments. The increasing adoption of heterogeneous and disaggregated architectures is also likely to shape future LLM systems. By decoupling compute, memory, and storage resources, these architectures enable more flexible and scalable resource allocation. This, in turn, can facilitate more efficient handling of large models and dynamic workloads.

Energy-aware computing is expected to become another major concern in LLM system design. Future systems must incorporate carbon-aware scheduling, energy-efficient model architectures, and runtime adaptations based on power constraints. Such approaches will be critical for ensuring the sustainability of large-scale AI deployments.

\subsection{\color{blue}Distributed Execution and System Software Challenges\color{black}}
\label{sec:system-software-challenges}

\subsubsection{Historical Context}

Traditional ML systems were typically developed and deployed as monolithic applications, tightly coupled with specific frameworks, execution environments, and hardware accelerators. These systems prioritized performance and simplicity over modularity, as workloads were relatively stable and predictable. With the increasing popularity of cloud computing, there is a paradigm shift toward microservices-based architectures, where applications are decomposed into loosely coupled services that can be independently developed, deployed, and scaled~\cite{microservices}. Technologies such as containerization and orchestration platforms, such as Docker Swarm and Kubernetes became foundational to modern distributed systems, enabling elasticity, fault tolerance, and portability~\cite{kubernetes}.

The integration of LLMs into these cloud-native ecosystems has introduced significant challenges. Unlike traditional services, LLM workloads exhibit tight coupling between computation, memory, and communication, along with highly dynamic execution patterns. While microservices architectures emphasize modularity and abstraction, LLM systems often require cross-layer optimization and fine-grained coordination. This fundamental mismatch has exposed limitations in existing system software abstractions, which were not designed to accommodate the scale, variability, and performance sensitivity of LLM workloads. Consequently, deploying and managing LLM-based services within cloud-native environments introduces new complexities that span the entire software stack.

\subsubsection{Open Research Questions}

\color{blue}
Based on the system and software challenges and limitations discussed above, we
identify the following open research questions for future cloud-native LLM systems:
\begin{itemize}
    \item RQ1: What new cloud-native abstractions are required for supporting LLM-native applications?
    \item RQ2: How can distributed runtime systems provide efficient and elastic execution for large-scale LLM services?
    \item RQ3: How can cloud-native orchestration frameworks become LLM-aware and self-adaptive?
\end{itemize}\color{black}

\subsubsection{State of the Art}

Current LLM systems are typically built on software stacks that emphasize modularity, portability, and scalability. Containerization technologies enable packaging of model dependencies, while orchestration frameworks such as Kubernetes provide mechanisms for auto-scaling, fault tolerance, and service discovery. Model serving frameworks further decompose inference pipelines into components responsible for request routing, model execution, caching, and monitoring. This modular design allows flexible deployment and integration with broader application ecosystems.

However, these abstractions largely treat LLM inference as a black-box service, limiting opportunities for system-level optimization. Recent research has highlighted the inefficiencies of this approach, particularly in multi-tenant environments where fine-grained scheduling and resource sharing are critical. Systems such as vLLM~\cite{vllm} and related inference engines introduce LLM-aware optimizations, including efficient memory management for key-value caches and continuous batching, significantly improving throughput and latency under dynamic workloads~\cite{flexgen}. Similarly, research on inference-serving systems has explored techniques for request-aware scheduling, prioritization, and latency-aware batching, demonstrating the importance of exposing workload semantics to the system layer~\cite{orca, mohammadi2025evaluation}.

Moreover, some studies focus on LLM-aware runtimes and serving stacks that integrate model execution with system-level scheduling. These systems aim to bridge the gap between model internals and infrastructure by enabling fine-grained control over execution pipelines. Recent efforts in disaggregated serving architectures separate compute and memory resources, allowing more flexible scaling and improved utilization~\cite{distserve}. Additionally, systems research has also explored unified serving layers that support multiple models and frameworks, reducing duplication and improving maintainability in production environments~\cite{chen2026towards}. Observability and performance analysis have also received increasing attention. Emerging tools provide fine-grained tracing and profiling capabilities across the LLM stack, capturing interactions between model execution, runtime scheduling, and infrastructure behavior. These tools are essential for diagnosing performance bottlenecks in complex deployments, though they remain limited in their ability to provide actionable insights across layers~\cite{watson}.

\subsubsection{Challenges and Limitations}

Despite these advances, the LLM software ecosystem remains fragmented. A key challenge in LLM system software lies in the mismatch between cloud-native abstractions and LLM workload characteristics. Microservices architectures, while flexible and scalable, introduce overheads in communication, serialization, and service coordination. These overheads can significantly impact performance for LLM workloads, which require low-latency and high-throughput execution with tight coupling between components. As a result, simple decomposition of LLM pipelines into microservices can lead to inefficiencies that negate the benefits of modularity.

Another major limitation is the lack of cross-layer visibility. Existing system software often operates without awareness of model-level behavior, such as token generation patterns, attention mechanisms, or memory access characteristics. This lack of visibility prevents effective coordination between layers, leading to suboptimal decisions in scheduling, caching, and resource allocation. Consequently, many optimizations remain confined to individual layers rather than being applied holistically across the system.

Debugging and performance diagnosis in LLM systems present additional challenges. The software stack spans multiple layers, including model code, runtime libraries, serving frameworks, orchestration platforms, and underlying infrastructure. Performance issues can arise from complex interactions between these layers, making root-cause analysis difficult. Existing observability tools are often insufficient for capturing these interactions, as they were not designed for the tightly coupled and dynamic nature of LLM workloads.

The rapid pace of innovation in LLMs further exacerbates system complexity. New models, architectures, and optimization techniques are introduced frequently, requiring continuous updates to system software. This leads to compatibility issues and challenges in maintaining stable production environments. Finally, heterogeneity across frameworks and hardware platforms introduces additional complexity. Supporting multiple backends while maintaining performance portability remains an open problem.

\subsubsection{Future Prospects}

Future system software for LLMs is likely to evolve toward LLM-aware, cross-layer, and adaptive architectures that address the limitations of current cloud-native approaches. One key direction is the development of abstractions that explicitly capture LLM workload semantics, enabling tighter integration between model execution and system-level decision-making. Such abstractions could expose information about token-level execution, memory usage, and computational dependencies, allowing systems to perform more informed scheduling and optimization.

Another key direction is the emergence of LLM-aware middleware and orchestration frameworks, which can automatically manage service decomposition, dependency resolution, and performance tuning, thus reducing the operational burden on developers. By incorporating knowledge of model behavior and workload dynamics, such frameworks could achieve better trade-offs between flexibility and performance compared to traditional microservices architectures.

Advances in observability and debugging tools will also play a critical role. Future systems may provide unified, cross-layer visibility into LLM execution, enabling more effective performance diagnosis and optimization. This could include integration of tracing, profiling, and anomaly detection capabilities tailored to LLM workloads, as well as automated root-cause analysis. Finally, the convergence of system software and machine learning techniques is likely to drive the next generation of LLM systems. Learning-based system optimization, where models are used to predict workload behavior and guide system decisions, represents a promising approach for managing complexity in dynamic environments. Such systems could continuously adapt to changing conditions, improving performance, efficiency, and robustness over time.

\subsection{\color{blue}Resource Management and Serving Challenges\color{black}}
\label{sec:operational-challenges}

\subsubsection{Historical Context}

Resource management in distributed systems has historically evolved in response to the dominant workload paradigms of each era. Early distributed computing environments were primarily designed to support enterprise applications such as transactional databases, web services, and business process platforms~\cite{brambilla2006process}. These workloads exhibited relatively stable and predictable resource consumption patterns, characterized by steady CPU and memory usage, moderate horizontal scalability, and strict latency and availability requirements. Consequently, resource management strategies emphasized isolation, fault tolerance, and elasticity, with scaling decisions driven by relatively simple workload indicators. As cloud computing matured, the introduction of virtualization and containerization technologies enabled more flexible and efficient resource utilization. Infrastructure tools such as Terraform and Ansible facilitated reproducible system configurations~\cite{jha2020challenges}. These systems operated under the assumption that workloads could be decomposed into interchangeable units, such as virtual machines, containers, or pods, which could be replicated or rescheduled with minimal overhead.

In parallel, batch processing systems and data analytics frameworks introduced scheduling policies optimized for throughput and fairness. These systems relied on queue-based scheduling, task packing, and speculative execution to maximize utilization in multi-tenant environments~\cite{zhang2016parallel}. Importantly, workloads were typically finite, loosely coupled, and tolerant to re-execution, enabling aggressive optimization strategies such as preemption. Stream processing systems extended these principles to continuous workloads, maintaining assumptions of incremental scalability and well-defined resource requirements~\cite{liu2020resource}.

The rise of ML and DL introduced new requirements, particularly with the need for accelerator-aware scheduling and gang scheduling for distributed training~\cite{ye2024deep}. However, these workloads remained largely phase-bounded and restartable, fitting within the broader assumptions of cloud resource management. In contrast, LLM workloads disrupt these established models. They are characterized by tightly coupled computation, high memory demands, long-running inference sessions, and significant variability in request patterns. These properties challenge fundamental assumptions of resource interchangeability, elasticity through replication, and coarse-grained monitoring, exposing the limitations of existing operational and resource management frameworks.

\subsubsection{Open Research Questions}

\color{blue}
Based on the resource management challenges and limitations
discussed above, we identify the following open research questions for efficient,
adaptive, and reliable LLM service management:

\begin{itemize}
    \item RQ1: How can resource management systems achieve SLO-aware optimization for multi-tenant LLM services?
    \item RQ2: How can LLM inference systems dynamically adapt resources to unpredictable workload variations?
    \item RQ3: How can prefill/decode execution characteristics be exploited for efficient LLM serving?
\end{itemize}\color{black}

\subsubsection{State of the Art}

Operational support for LLM systems has largely adapted techniques from cloud resource management, including autoscaling~\cite{StatuScale}, load balancing, and admission control. These mechanisms are effective for handling steady-state workloads but are less effective under the bursty and highly variable access patterns typical of LLM services. Recent work has demonstrated that simple autoscaling policies, which rely on coarse-grained metrics such as CPU or GPU utilization, often fail to capture the true resource demands of LLM inference, leading to either overprovisioning or service degradation~\cite{heisler2025llm}.

Dynamic batching has emerged as a central technique for improving accelerator utilization. By aggregating requests based on arrival patterns and latency constraints, systems can significantly increase throughput, particularly for GPU-based inference. However, batching introduces complex trade-offs between latency and efficiency, requiring careful tuning of batch sizes and scheduling policies. Existing serving frameworks incorporate request-aware scheduling and iteration-level execution to improve responsiveness under dynamic workloads~\cite{zhu2025nanoflow, ye2025flashinfer}. Similarly, advances in memory management and scheduling, such as those introduced in Jenga~\cite{zhang2025jenga}, enable more efficient utilization of GPU memory resources, which are often the primary bottleneck in LLM serving.

Predictive resource management has also gained attention as a means to address workload variability. By leveraging historical workload traces and lightweight forecasting models, systems can anticipate demand fluctuations and proactively allocate resources. Recent approaches integrate ML-based predictors with autoscaling policies, enabling more responsive and efficient resource provisioning~\cite{throttlelm}. In addition, hybrid deployment models that span cloud and edge environments aim to reduce latency and network overhead by placing inference closer to end users. While promising, these approaches introduce new challenges in coordination, consistency, and resource fragmentation~\cite{li2025collaborative}.

\subsubsection{Challenges and Limitations}

As autoscaling and load balancing decisions are triggered by observed system metrics, they key limitation is the reliance on reactive control mechanisms which may lag behind actual workload changes. Since the inference demand can spike rapidly and unpredictably, any delays can result in significant performance degradation or excessive overprovisioning. The inability to anticipate workload dynamics limits the effectiveness of existing operational policies.

Another key challenge lies in the granularity of resource management. Traditional systems operate at the level of virtual machines or containers, abstracting away finer-grained resources such as GPU memory, interconnect bandwidth, and cache state. However, LLM workloads are highly sensitive to these resources, particularly memory and communication bandwidth. The lack of mechanisms to explicitly manage and schedule these fine-grained resources leads to inefficiencies and contention, especially in multi-tenant environments.

Fairness and isolation also present significant challenges. In shared LLM platforms, workloads with different characteristics and priorities compete for limited resources. Ensuring that no single workload monopolizes resources while maintaining high utilization is a non-trivial problem. Existing fairness mechanisms, which are often designed for homogeneous workloads, do not adequately capture the heterogeneity and performance sensitivity of LLM services.

Energy efficiency is another critical issue. Current resource management systems rarely incorporate energy or carbon considerations into their decision-making processes resulting in energy-inefficient regimes, particularly during periods of low utilization or when overprovisioning resources to meet peak demand. The absence of energy-aware scheduling policies limits the sustainability of large-scale LLM deployments.

\subsubsection{Future Prospects}

Future resource management systems for LLMs need to adopt predictive, fine-grained, and multi-objective optimization strategies that move beyond the limitations of current approaches. Integration of workload forecasting and user behavior modeling into resource allocation decisions can be the main focus. By leveraging historical data and real-time telemetry, systems can anticipate demand fluctuations and proactively provision resources, reducing both latency and cost.

Another promising avenue is the development of fine-grained resource abstractions that expose critical components such as GPU memory, cache state, and communication bandwidth to the scheduler. This would enable more precise control over resource allocation and improve utilization in heterogeneous and multi-tenant environments. Coupled with advances in disaggregated architectures, such abstractions could facilitate more flexible and efficient resource sharing.

Multi-tenant platforms will increasingly require sophisticated fairness and isolation mechanisms that account for workload heterogeneity and service-level objectives. This may involve the use of priority-aware scheduling, admission control policies, and workload classification techniques to ensure equitable resource distribution while maintaining high system efficiency.

Energy-aware resource management is also expected to become a central concern. Future systems may incorporate carbon-aware scheduling, dynamic power management, and energy-efficient workload placement strategies. These approaches will be critical for reducing the environmental impact of large-scale LLM deployments while maintaining performance guarantees.

\subsection{Security and Privacy Challenges}
\label{sec:privacy-security-challenges}

\subsubsection{Historical Context}

Privacy and security have long been critical concerns in distributed systems and cloud computing environments. Traditional enterprise applications typically rely on well-defined access control mechanisms, secure storage architectures, and network isolation policies to protect sensitive information. Earlier machine learning models were generally trained on relatively constrained datasets within controlled environments, which reduced the risk of large-scale data exposure.

The emergence of Large Language Models fundamentally changes this landscape. LLMs are trained on extremely large and heterogeneous datasets aggregated from web-scale sources, proprietary knowledge bases, and user-generated content. These datasets may inadvertently contain personally identifiable information (PII), confidential business data, or copyrighted material. Recent studies have demonstrated that large language models can memorize segments of their training data and reproduce them during inference under certain conditions, raising significant privacy concerns~\cite{carlini2021extracting,carlini2022quantifying}. Further research shows that large-scale generative models may unintentionally reveal sensitive information through prompt-based extraction or model inversion techniques, highlighting the importance of stronger privacy protections during training and deployment.

In addition, LLM services are increasingly deployed as multi-tenant cloud platforms, where models interact dynamically with user queries, external APIs, and retrieval systems. These architectures introduce new attack surfaces across distributed components, including model-serving interfaces, orchestration platforms, vector databases, and inter-service communication channels. As LLM-based services become widely integrated into enterprise systems, healthcare platforms, and public-sector applications, ensuring robust privacy and security protections across the entire system stack becomes essential.

\subsubsection{Open Research Questions}

\color{blue}

Based on the privacy and security challenges and limitations discussed above, we
identify the following open research questions for secure and trustworthy
cloud-native LLM systems:

\begin{itemize}
    \item RQ1: How can personalized LLM services be enabled while preserving data privacy and security?
    \item RQ2: How can distributed LLM systems ensure trustworthy and secure operation in multi-tenant environments?
    \item RQ3: How can access control, authentication, and auditing mechanisms be enforced consistently across multi-tenant LLM platforms?
\end{itemize}\color{black}

Addressing these questions requires coordinated solutions across multiple layers of the LLM ecosystem, including model training frameworks, distributed system architectures, and cloud infrastructure platforms.

\subsubsection{State of the Art}

Recent research has explored several approaches to enhance privacy and security in large-scale machine learning systems and LLM platforms.

One widely studied direction focuses on privacy-preserving machine learning techniques. Differential privacy introduces calibrated noise during model training to ensure that individual data points cannot be inferred from model outputs~\cite{abadi2016deep}. This technique provides formal privacy guarantees and has been applied in large-scale deep learning systems. Another important approach is federated learning, which allows models to be trained collaboratively across distributed data sources while keeping raw data localized on edge devices or institutional servers~\cite{mcmahan2017communication}. This approach reduces the risk of centralized data leakage and is particularly relevant for sensitive domains such as healthcare and finance.

Another promising research direction involves secure computation frameworks. Trusted Execution Environments (TEEs), such as Intel SGX, allow sensitive computations to be executed within hardware-protected enclaves that isolate both data and model parameters from the host operating system~\cite{ohrimenko2016oblivious}. These technologies have been explored for secure machine learning inference in cloud environments where sensitive data must be protected from infrastructure-level threats.

More advanced cryptographic techniques such as homomorphic encryption (HE) and secure multi-party computation (SMC) enable machine learning models to perform computations on encrypted data without revealing the underlying inputs. While these techniques offer strong privacy guarantees, they currently introduce substantial computational overhead, which limits their practicality for large-scale LLM deployments.

From a systems perspective, modern cloud-native infrastructures incorporate defense-in-depth security mechanisms, including encryption of data in transit and at rest, identity and access management frameworks, container isolation, and runtime monitoring. These mechanisms are increasingly integrated into LLM serving infrastructures deployed on container orchestration platforms such as Kubernetes~\cite{kubernetes}.

In addition to privacy-preserving training techniques, researchers have recently identified new classes of security threats specific to LLM systems. Prompt injection attacks exploit the natural language interface of LLMs to manipulate system behavior or override safety constraints~\cite{greshake2023not}. Similarly, model extraction attacks attempt to reconstruct proprietary models by repeatedly querying prediction APIs~\cite{tramer2016stealing}. Recent work also shows that adversarial prompts and jailbreak techniques can circumvent alignment safeguards and force models to generate restricted content~\cite{zou2023universal}. These emerging threats highlight the need for stronger security mechanisms that protect not only the model itself but also the broader LLM service infrastructure.

\subsubsection{Challenges and Limitations}

Despite significant progress, several challenges limit the effectiveness of current privacy and security approaches for LLM systems.

First, many privacy-preserving techniques introduce substantial computational overhead. Methods such as differential privacy, homomorphic encryption, and secure multi-party computation can significantly increase training and inference costs. For LLMs with billions of parameters, these overheads can become prohibitive in large-scale deployments.

Second, secure execution environments face scalability limitations. Hardware enclaves often have restricted memory capacity and limited compatibility with GPU accelerators, making them difficult to integrate with modern LLM pipelines that rely heavily on high-performance hardware.

Third, ensuring end-to-end security across distributed LLM architectures remains a major challenge. LLM systems often involve multiple interconnected services, including data preprocessing pipelines, retrieval components, vector databases, orchestration layers, and external APIs. Each component introduces potential vulnerabilities that must be addressed through coordinated security policies.

Finally, LLM systems are inherently interactive and probabilistic, which creates new security challenges not present in traditional software systems. Recent studies indicate that LLM-based platforms remain vulnerable to coordinated attacks that exploit both the model and surrounding infrastructure components, including retrieval systems and tool integrations~\cite{das2025security}. These vulnerabilities highlight the need for holistic security frameworks that consider the entire LLM service architecture rather than focusing solely on model-level defenses.

\subsubsection{Future Prospects}

Future research must address privacy and security challenges through tighter integration between machine learning models and distributed system infrastructures.

One promising direction involves extending cloud-native security primitives such as identity-aware service meshes, confidential computing frameworks, and policy-driven access control systems to support LLM-specific workloads. These mechanisms can enforce strong security guarantees while maintaining the scalability and flexibility required for distributed AI systems.

Another emerging direction is privacy-aware resource orchestration, where system schedulers incorporate privacy and security requirements into deployment decisions. For example, sensitive workloads may be automatically scheduled on confidential computing nodes or isolated clusters with stronger hardware-based security protections.

In addition, advances in privacy-preserving inference techniques, including optimized homomorphic encryption and hardware-accelerated cryptographic primitives, may eventually enable secure model execution without exposing sensitive data.

Finally, defending against LLM-specific threats such as prompt injection, adversarial prompts, and model extraction attacks will require a combination of model-level safeguards, runtime monitoring, and system-level policy enforcement mechanisms. Integrating anomaly detection systems and automated auditing pipelines into LLM serving infrastructures will be essential for ensuring the reliability and trustworthiness of next-generation AI platforms.

As LLM-based services continue to expand across cloud and edge environments, privacy and security must be treated as first-class design objectives rather than optional add-ons. Achieving this goal will require interdisciplinary collaboration across machine learning, distributed systems, and cybersecurity research communities.

\subsection{Data Challenges}
\label{sec:data-challenges}

\subsubsection{Historical Context}

Data management has long been a foundational concern in distributed systems and ML pipelines. Traditional data-intensive systems were designed around relatively static datasets, often curated offline and processed through well-defined pipelines~\cite{bian2022machine}. In early ML/DL workflows, datasets were versioned infrequently, stored in centralized or distributed file systems, and accessed in bulk during training. These pipelines emphasized reproducibility and consistency, with limited need for real-time updates or continuous data integration.

As data systems evolved, the emergence of large-scale analytics and streaming platforms introduced mechanisms for handling dynamic and continuously generated data. Distributed storage systems, data lakes, and stream processing frameworks enabled scalable ingestion, processing, and querying of heterogeneous data sources~\cite{ait2023spatial, hai2023data}. However, these systems still largely assumed a separation between data processing and model execution, with clearly defined boundaries between offline training and online inference.

The rise of LLMs fundamentally alters these assumptions as they depend on massive, heterogeneous, and continuously evolving datasets, both during training and at inference time. In particular, the increasing adoption of retrieval-augmented generation (RAG) blurs the boundary between data management and model execution, requiring systems to dynamically access, filter, and integrate external knowledge sources during inference~\cite{zhao2024retrieval}. This shift transforms data from a static input into an active, continuously evolving component of the system, tightly coupled with runtime behavior and performance.

\subsubsection{Open Research Questions}

\color{blue}
Based on the data management challenges and limitations discussed above, we
identify the following open research questions for scalable, efficient, and
trustworthy data management throughout the LLM lifecycle:

\begin{itemize}
    \item RQ1: How can cloud-native infrastructures efficiently manage distributed LLM data and model states?
    \item RQ2: How can retrieval-augmented LLMs efficiently integrate external data sources?
    \item RQ3: How can data governance and compliance be enforced across distributed LLM pipelines?
\end{itemize}\color{black}

\subsubsection{State of the Art}

Modern LLM systems rely on large-scale distributed storage infrastructures and data pipelines to manage both training datasets and inference-time knowledge sources. High-performance storage systems, including distributed object stores and parallel file systems, are commonly used to support large-scale data access. To mitigate latency and bandwidth constraints, systems increasingly incorporate multi-level caching and data prefetching strategies, particularly for frequently accessed data during training and inference~\cite{sim2022computational,gu2023high}. These optimizations are essential for maintaining high throughput in data-intensive workloads, where data movement can otherwise dominate system performance.

A significant development in recent years is the adoption of retrieval-augmented generation (RAG) architectures, which integrate external data sources into the inference process. These systems typically combine vector databases, dense retrieval models, and LLM inference engines to enable dynamic knowledge access~\cite{gao2023retrieval}. Efficient vector search and indexing mechanisms have become critical components of these systems, with ongoing research focusing on scalable approximate nearest neighbor search and distributed indexing techniques~\cite{echihabi2021new,yin2025unleash}. Moreover, there is growing interest in data-aware system optimization, where scheduling and execution decisions are informed by data placement and access patterns. Recent studies explored co-designing data pipelines and model execution to minimize data movement and improve locality~\cite{li2025hydraulis}. Similarly, caching strategies tailored to LLM workloads, such as semantic caching and query-aware caching, have been proposed to reduce redundant computation and data access~\cite{iyengar2025generative}. Data versioning and provenance tracking are also receiving increased attention, particularly in the context of reproducibility and compliance~\cite{hohensinner2026tracing}. However, these mechanisms are often limited in scalability and integration with real-time data pipelines.

\subsubsection{Challenges and Limitations}

One of the most significant challenges in LLM data management is the cost and complexity of data movement. As datasets grow in size and distribution, transferring data between storage, compute nodes, and inference services becomes a major bottleneck.  This is particularly significant in RAG systems, where real-time retrieval introduces additional latency and network overhead. Existing systems often lack mechanisms to effectively minimize or hide these costs, leading to inefficiencies in both training and inference. Ensuring data quality, freshness, and consistency presents another major challenge. LLMs increasingly rely on external and user-generated data sources, which may be noisy, inconsistent, or outdated. Maintaining high-quality data pipelines requires continuous validation, filtering, and updating, which can be difficult to achieve at scale. Moreover, different components of the system may operate on different versions of the data, leading to inconsistencies that affect model behavior and reliability.

Data governance and compliance further complicate system design. Regulations such as data protection laws impose strict requirements on how data is stored, processed, and accessed. In distributed LLM pipelines, enforcing these requirements across multiple components and geographic regions is non-trivial. Ensuring that sensitive data is handled appropriately, while maintaining system performance, requires sophisticated access control, auditing, and encryption mechanisms. 

Another limitation is the lack of unified abstractions for data management across the LLM lifecycle. Training, fine-tuning, and inference often rely on different data systems and formats, leading to fragmentation and increased operational complexity. This fragmentation impacts reproducibility, as it becomes difficult to track how data flows through the system and influences model outputs. Finally, balancing trade-offs between consistency, latency, and scalability remains an open problem. Strong consistency guarantees can introduce significant overhead, while weaker consistency models may lead to stale or inconsistent results. Designing systems that can adaptively balance these trade-offs based on application requirements is a key challenge for future research.

\subsubsection{Future Prospects}

Future data management systems for LLMs will be data-centric, tightly integrated architectures that unify storage, retrieval, and computation. One promising direction is the development of systems that explicitly optimize data locality, placing data closer to computation or dynamically moving computation closer to data. Such approaches can significantly reduce data movement costs and improve overall system efficiency. 

Another important direction is the integration of intelligent caching and retrieval mechanisms. By leveraging semantic understanding of queries and data, systems can proactively cache relevant information and reduce redundant retrieval operations. This is particularly relevant for RAG systems, where efficient access to external knowledge sources is critical for performance. Advances in data versioning and provenance tracking are expected to improve reproducibility and accountability in LLM systems. 

Future platforms must provide built-in support for tracking data lineage across the entire model lifecycle, enabling fine-grained auditing and easier compliance with regulatory requirements. These capabilities will be essential as LLMs are increasingly deployed in sensitive and regulated domains. 

Data governance will also become more tightly integrated with system design. Rather than being treated as an external concern, governance mechanisms such as access control, privacy enforcement, and compliance monitoring may be embedded directly into data pipelines and system runtimes. This integration can help ensure that data is handled responsibly without compromising performance.

Finally, the convergence of data management and ML is likely to drive the development of self-optimizing data systems. By leveraging machine learning techniques, these systems can automatically adapt data placement, caching strategies, and retrieval policies based on observed workload patterns. This shift toward intelligent, adaptive data management will be critical for supporting the scale, complexity, and dynamism of future LLM applications.

\subsection{\color{blue}Standardization Challenges\color{black}}
\label{sec:standardization}

\subsubsection{Historical Context}

Standardization has been a foundational enabler in the evolution of distributed systems, providing the common abstractions and interfaces necessary for interoperability, portability, and large-scale ecosystem development. Widely adopted standards have allowed heterogeneous systems to communicate seamlessly, reduced integration complexity, and accelerated innovation by decoupling system components. As distributed computing matured, standardization efforts helped transform fragmented technological landscapes into cohesive platforms, enabling the emergence of robust multi-vendor ecosystems and scalable services. 

In contrast, the rapidly evolving landscape of LLM systems has yet to reach a comparable level of convergence. The pace of advancement in models, hardware accelerators, and software frameworks has led to a proliferation of competing approaches, each introducing its own abstractions, interfaces, and deployment paradigms. While this diversity has fueled rapid progress, it has also resulted in significant fragmentation, making it increasingly difficult to achieve interoperability and portability across systems. This divergence highlights the growing importance of establishing meaningful standards for LLM systems.

\subsubsection{Open Research Questions}
\color{blue}
Based on the standardization challenges and limitations discussed above, we
identify the following open research questions for interoperable, portable, and
reproducible LLM systems:

\begin{itemize}
    \item RQ1: What standards are needed to enable interoperable LLM deployment across heterogeneous infrastructures?
    \item RQ2: How can cloud-native ecosystems provide unified programming and management interfaces for diverse LLM workloads?
    \item RQ3: How can benchmarking methodologies accurately evaluate cloud-native LLM systems?
\end{itemize}\color{black}

\subsubsection{State of the Art}

Recent efforts toward standardization in LLM systems have focused on defining common formats, interfaces, and deployment abstractions. Model representation formats such as ONNX and MLIR-based approaches aim to provide a hardware-agnostic intermediate representation, enabling models to be executed across diverse platforms with minimal modification~\cite{agostini2022mlir,majumder2023hir}. These efforts are complemented by standardized inference APIs and serving interfaces, which seek to unify how models are accessed and deployed across different environments. Open-source frameworks have played a particularly influential role in shaping de facto standards. Systems such as Hugging Face Transformers~\footnote{\href{https://www.google.com/url?q=https://huggingface.co/&sa=D&source=docs&ust=1774321803233395&usg=AOvVaw2LnjASkv7bvTDvDgfH44bx}{https://huggingface.co/}}, Ollama ~\footnote{https://ollama.com/} and LM Studio~\footnote{https://lmstudio.ai/} have established widely adopted conventions for model packaging, tokenization, and inference workflows for developers. However, these conventions often evolve organically rather than through formal standardization processes, leading to inconsistencies and compatibility challenges across ecosystems. 

In the deployment context, cloud-native technologies are beginning to incorporate LLM-specific abstractions. For example, Kubernetes-based extensions and custom resource definitions are being used to describe model deployments, resource requirements, and scaling policies~\cite{masood2022machine}. While these approaches improve integration with existing infrastructure, they remain largely platform-specific and lack cross-provider standardization. Similarly, efforts in disaggregated and multi-tenant serving systems highlight the need for standardized resource descriptors and scheduling interfaces that can capture the unique characteristics of LLM workloads~\cite{distserve}.

Benchmarking and evaluation represent another area where standardization is still emerging. Recent work has proposed more realistic and comprehensive benchmarks for LLM inference and serving, emphasizing metrics such as latency, throughput, and cost under dynamic workloads~\cite{lazuka2024llm,jegham2025hungry}. However, there is still no widely accepted standard for evaluating system-level performance of LLM deployments, making it difficult to compare solutions and reproduce results.

\subsubsection{Challenges and Limitations}

A central challenge in standardizing LLM systems is the tension between rapid innovation and the need for stable interfaces. The field is evolving at an unprecedented pace, with frequent advances in model architectures, training techniques, and hardware accelerators. Establishing standards too early risks constraining innovation by locking systems into suboptimal abstractions. Conversely, the absence of standards leads to fragmentation, making it difficult to build interoperable and sustainable systems.

Another significant limitation is the diversity of the LLM ecosystem. Systems span multiple layers, including model architectures, runtime frameworks, serving infrastructures, and hardware platforms. Each layer has distinct requirements and evolves at different rates, complicating the design of unified standards. As a result, many existing standardization efforts focus on narrow aspects of the stack, such as model formats or APIs, without addressing cross-layer integration. Fine-tuning methods including in-context learning and low-rank adaptation (LoRA) can make the benchmarking even more complicated~\cite{wu2025survey}.

Benchmarking further illustrates the limitations of current approaches. Existing benchmarks often fail to capture the complexity of real-world LLM workloads, such as multi-tenant environments, dynamic request patterns, and heterogeneous hardware configurations. This lack of representative evaluation frameworks makes it difficult to assess the effectiveness of different systems and hinders the development of best practices. Finally, governance and coordination present additional challenges. Standardization requires collaboration across academia, industry, and open-source communities, each with different incentives and priorities. Achieving consensus on common standards is a complex and time-consuming process, particularly in a rapidly evolving field.

\subsubsection{Future Prospects}

Standardization is likely to become a key enabler of scalable, interoperable, and sustainable LLM ecosystems. One important direction is the development of layered standards, where different levels of abstraction, including model representation, runtime interfaces, and deployment specifications, are standardized independently while maintaining clear interfaces between them. This approach can provide stability without limiting innovation at individual layers. Standardized deployment models, resource descriptors, and orchestration interfaces could enable seamless integration of LLMs into multi-cloud and hybrid environments. Such standards would reduce vendor lock-in and facilitate portability across platforms.

Another critical area is benchmarking and evaluation. Developing standardized benchmarks that reflect real-world workloads, including dynamic and multi-tenant scenarios, will be essential for advancing the field. These benchmarks can provide a common basis for comparison, enabling more rigorous evaluation of system designs and fostering reproducibility. This requires collaborations between academia and industry to shape effective standards. While academia can contribute principled abstractions and evaluation methodologies, industry can provide practical insights and drive adoption at scale. 

Finally, standardization efforts can increasingly incorporate automation and adaptability. Rather than fixed specifications, future standards could include mechanisms for extensibility and evolution, allowing systems to adapt to new models, hardware, and workloads. In this vision, cloud-native systems serve not only as execution platforms but also as unifying layers that enable collaborative, secure, and efficient development of LLM-based applications at scale.\\


\color{blue}
\subsection{Cross-cutting Synthesis of System Challenges}

\begin{table*}[t]
	\centering
	\caption{\color{blue}Cross-cutting relationships among the major system challenges for cloud-native LLM systems.}
	\label{tab:cross_cutting_challenges}
	{\color{blue}
	\renewcommand{\arraystretch}{1.15}
	\fontsize{8.5pt}{10pt}\selectfont
	\begin{tabularx}{\textwidth}{p{3.1cm}p{5.0cm}X}
		\toprule
		\textbf{Challenge Dimension} &
		\textbf{Key Cross-cutting Dependency} &
		\textbf{System-level Implication} \\
		\midrule
		
		Computational 
		&
		Hardware heterogeneity; communication; resource management
		&
		Topology-aware and accelerator-aware scheduling
		\\
		
		Distributed Execution and System Software 
		&
		Resource management; data and cache management
		&
		Cross-layer observability and LLM-aware runtime abstractions
		\\
		
		Resource Management and Serving 
		&
		Computation; hardware; QoS
		&
		Workload-aware, predictive, and multi-objective orchestration
		\\
		
		Security and Privacy 

		&
		Data management; distributed execution
		&
		End-to-end security, access control, and compliance
		\\
		
		Data 

		&
		Computation; communication; security
		&
		Data-aware placement, locality optimization, and governance
		\\
		
		Standardization
		&
		All challenge dimensions
		&
		Interoperable abstractions, interfaces, and deployment mechanisms
		\\
		
		\bottomrule
	\end{tabularx}
}
\end{table*}

The six challenge dimensions discussed above are closely interconnected rather
than independent. The computational characteristics of LLMs influence system
software abstractions and resource management policies, while data movement,
privacy requirements, and standardization further constrain the design of the
overall infrastructure. Therefore, addressing LLM system challenges requires
cross-layer coordination rather than isolated optimization within individual
system components. Table~\ref{tab:cross_cutting_challenges} summarizes the major
cross-cutting relationships among the six challenge dimensions and their
corresponding system-level implications.

Several important cross-cutting relationships emerge from the preceding analysis.
As summarized in Table~\ref{tab:cross_cutting_challenges}, \textit{hardware
	heterogeneity} directly connects computational challenges with operational and
resource management challenges. Differences in accelerator capabilities, memory
capacities, and interconnect characteristics affect workload placement,
parallelization strategies, scheduling decisions, and resource utilization.
Consequently, future resource managers need to jointly consider LLM execution
characteristics and hardware topology.

Second, \textit{cross-layer visibility} links system and software challenges with
resource management and data management. Limited visibility into model execution,
memory access patterns, intermediate states, and runtime behavior can result in
suboptimal scheduling, caching, and resource-allocation decisions. As reflected in
Table~\ref{tab:cross_cutting_challenges}, this motivates the development of unified
observability and runtime abstractions that expose LLM-relevant information across
system layers.

Third, \textit{data movement and locality} create a direct connection between data
management and computational and operational efficiency. The transfer of model
states, KV caches, training data, and retrieved knowledge across storage, memory,
accelerators, and network boundaries can introduce substantial communication
overhead and latency. Therefore, data-aware scheduling and computation-data
co-optimization are important for improving both performance and resource
utilization.

Fourth, \textit{privacy, security, and data governance} are inherently coupled.
Distributed LLM pipelines may involve user data, external knowledge sources,
retrieval systems, vector databases, and multiple service components. Access
control, auditing, encryption, compliance, and secure data processing therefore
need to be coordinated across the complete service pipeline rather than enforced
independently at individual components.

Finally, \textit{standardization} is a cross-cutting concern across all the
challenge dimensions. As shown in Table~\ref{tab:cross_cutting_challenges},
interoperable model representations, resource descriptors, runtime interfaces,
deployment specifications, and evaluation methodologies are needed to connect
heterogeneous hardware, software, resource-management, data, and security
mechanisms. Standardization can therefore serve as an enabling layer for
portability and coordinated optimization across the LLM infrastructure stack.

Overall, these relationships suggest that future cloud-native LLM systems should
evolve toward cross-layer, structure-aware, and jointly optimized infrastructures.
Rather than treating computation, software, resources, data, security, and
standardization as separate concerns, system designers need abstractions and
control mechanisms that explicitly capture their interactions. These
cross-cutting dependencies also motivate the resource management and optimization
mechanisms discussed in the next section, where computation, hardware, workload,
and service-level characteristics are jointly considered.

These cross-cutting dependencies motivate the resource management and optimization
mechanisms discussed in the next section, where computation, hardware, workload,
and service-level characteristics are jointly considered.

In addition, Table~\ref{tab:challenge_synthesis} provides a consolidated view of the six
challenge dimensions discussed in this section, their primary research focus,
the associated open research questions, and their cross-cutting connections with
other areas. The table emphasizes that these challenges are interdependent and
therefore cannot always be effectively addressed through isolated optimization
within individual system layers.

\begin{table*}[t]
	\centering
	\caption{\color{blue}Synthesis of the major challenges, open research questions, and
		cross-cutting relationships in cloud-native LLM systems.}
	\label{tab:challenge_synthesis}
	{\color{blue}
	\renewcommand{\arraystretch}{1.15}
	\footnotesize
	\begin{tabularx}{\textwidth}{
			p{2.7cm}
			p{2.4cm}
			p{6.0cm}
			X}
		\toprule
		\textbf{Challenge} &
		\textbf{Primary Area} &
		\textbf{Related Open Research Questions} &
		\textbf{Connections with Other Areas} \\
		\midrule
		
		Computational
		&
		Computation, heterogeneous accelerators, communication
		&
		Adaptive computation partitioning and scheduling; adaptive computation
		strategies; computation--communication co-optimization
		&
		Resource management; system and software; data management
		\\
		
		Distributed execution and system software
		&
		Runtime, orchestration, observability, cloud-native abstractions
		&
		LLM-aware system abstractions; elastic distributed runtimes; cross-layer
		observability and intelligent orchestration
		&
		Computation; resource management; data and cache management
		\\
		
		Resource management and serving
		&
		Scheduling, autoscaling, QoS, multi-tenancy
		&
		SLO-aware resource management; predictive autoscaling; fine-grained resource
		allocation; fairness and isolation
		&
		Computation; hardware heterogeneity; system and software
		\\
		
		Security and privacy
		&
		Secure execution, access control, privacy, threat mitigation
		&
		Privacy-preserving training and inference; secure multi-tenant execution;
		end-to-end protection and runtime security
		&
		Data management; system and software; resource placement
		\\
		
		Data
		&
		Data locality, retrieval, caching, provenance, governance
		&
		Data-aware execution; scalable versioning and provenance; efficient RAG
		integration; lifecycle-wide data management
		&
		Computation; communication; privacy and security
		\\
		
		Standardization
		&
		Interfaces, abstractions, interoperability, benchmarking
		&
		Standardized resource abstractions; portable interfaces; interoperable
		deployment; reproducible system-level benchmarking
		&
		All challenge areas
		\\
		
		\bottomrule
	\end{tabularx}
}
\end{table*}

As summarized in Table~\ref{tab:challenge_synthesis}, several system-level
concerns, including hardware heterogeneity, cross-layer visibility, data
locality, privacy, and standardization, act as bridges across multiple challenge
dimensions. These cross-cutting dependencies motivate integrated system designs
that jointly consider computation, resources, software, data, and security rather
than treating them as independent optimization objectives.

\color{black}
   \section{Resource Management and Optimization for LLMs}
\label{ch:4}


\color{blue}
The resource management and optimization techniques discussed in this section
should be viewed as direct responses to the challenges identified in Section~3.
The computational and communication bottlenecks discussed in Section~3.1
motivate heterogeneous- and topology-aware scheduling; the operational
limitations and reactive control mechanisms discussed in Section~3.3 motivate
elastic, predictive, and proactive resource management; the energy-efficiency
challenges motivate energy- and carbon-aware placement; and the multi-tenant
fairness and isolation challenges motivate adaptive QoS mechanisms. Likewise,
the cross-layer visibility limitations identified in Section~3.2 motivate the
observability-aware and AI-driven orchestration mechanisms discussed in
Section~4.4. Thus, Section~4 organizes representative techniques according to
the system problems they address, providing a solution-oriented complement to
the challenge-oriented analysis in Section~3. As shown in Fig.~3, this \color{blue}section organizes the main resource-management
mechanisms for LLM systems into four categories: elastic scheduling for heterogeneous hardware, energy- and carbon-aware placement, adaptive QoS in multi-tenant environments, and RL/AI-driven orchestration. \color{black}
\color{black}

\begin{figure}
    \centering
    \includegraphics[width=0.9\linewidth]{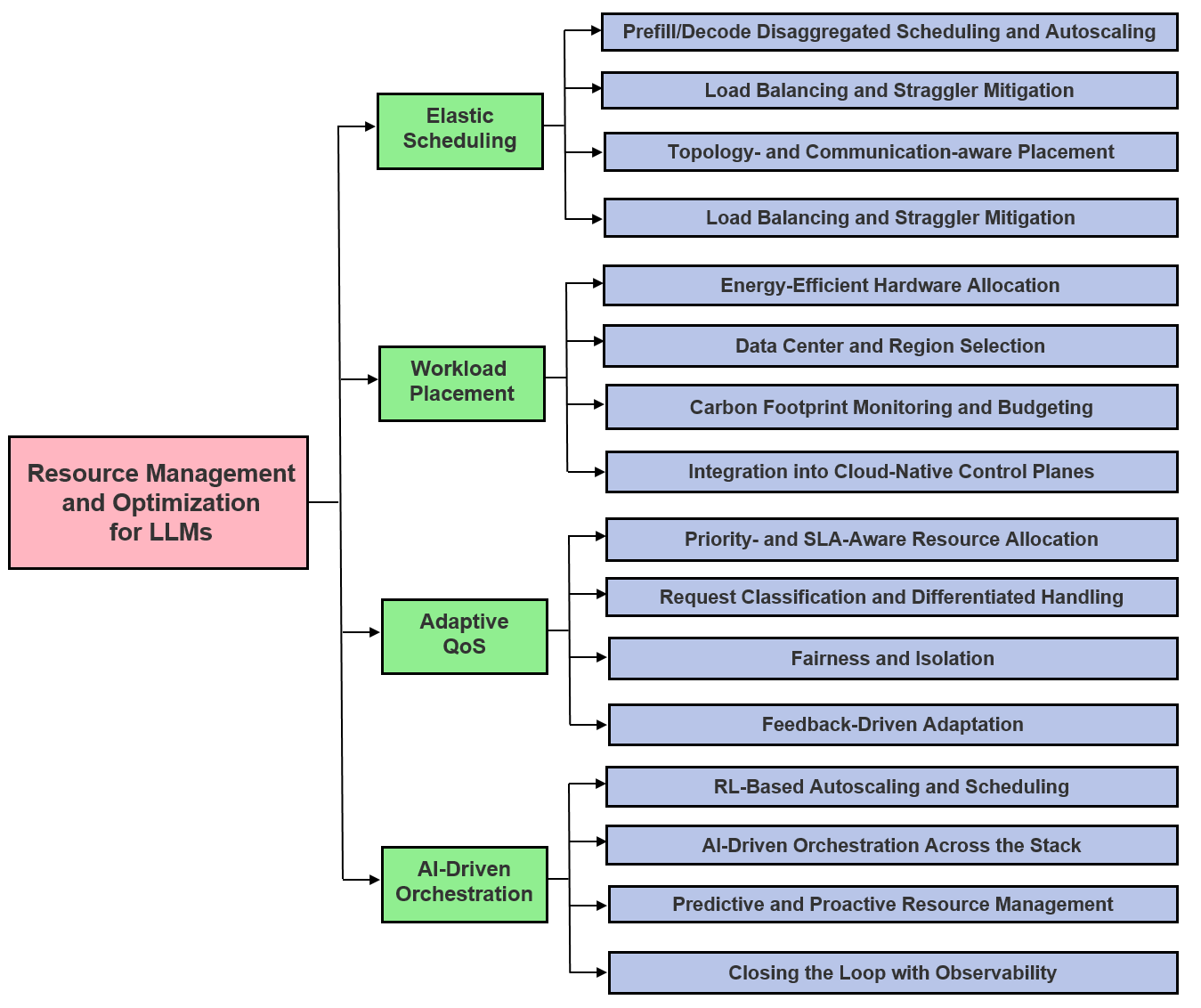}
    \caption{\color{blue}Resource management and optimization techniques for LLMs}
    \label{fig:RM}
\end{figure}

\subsection{Elastic Scheduling Policies for Heterogeneous Hardware}
\label{sec:elastic-scheduling}

Elastic scheduling for LLM serving arises from the interaction between highly dynamic request patterns and increasingly heterogeneous accelerator clusters~\cite{BrownoutServe}. Unlike traditional microservices, where CPU utilization and request rate often provide sufficient scheduling signals, LLM systems must respond to variability in prompt length, generation length, model structure, adapter activation, expert routing, and context size. At the same time, the hardware substrate is no longer a uniform GPU pool. Real deployments span different accelerator generations, memory capacities, interconnect fabrics, and auxiliary devices such as CPUs or lower-end GPUs. The objective of elastic scheduling is therefore to continuously match different workload components to the most suitable resources while preserving responsiveness under bursty demand. 


\subsubsection{Prefill/Decode Disaggregated Scheduling and Autoscaling}
\label{sec:pddisaggregation}

Prefill/decode disaggregation is motivated by a simple systems observation: LLM inference is not a uniform execution process, but a sequence of phases with different bottlenecks. The prefill phase is dominated by dense matrix computation and benefits from high compute throughput and strong parallelism. The decode phase is more constrained by KV-cache access, memory capacity, and memory bandwidth because generation proceeds token by token. \color{blue} These mechanisms directly address the hardware heterogeneity, fine-grained
scheduling, and communication overhead challenges identified in
Sections~3.1 and~3.3. In particular, separating compute-intensive prefill from
memory-intensive decode provides an opportunity to match different execution
phases with heterogeneous resources and to reduce contention across the serving
pipeline\color{black}. DistServe explicitly separates prefill and decode to optimize TTFT and TPOT under latency constraints \cite{distserve}\cite{DOPD} . ShuffleInfer further combines prompt chunking, P/D disaggregation, and two-level scheduling for mixed downstream workloads \cite{R5}. Splitwise extends phase splitting to heterogeneous clusters and shows that compute-intensive and memory-intensive phases can be mapped to different machines under throughput, cost, and power considerations \cite{R4}. FastServe addresses latency from another angle by introducing token-level preemptive scheduling to reduce head-of-line blocking in interactive serving \cite{R6}. 

The advantages of phase disaggregation must be considered together with its costs. P/D disaggregation introduces state transfer, tighter coupling between independently scaled pools, and additional routing complexity. If interconnect bandwidth is limited, prompts are short, or KV states are fragmented, migration overhead can offset the gain and even worsen tail latency. This is why recent systems increasingly emphasize bandwidth-aware placement, overlap of transfer and computation, and comparison against monolithic or chunked-prefill baselines rather than assuming that disaggregation is always beneficial \cite{distserve} \cite{R4}\cite{R5}. 

Once serving is decomposed into specialized resource pools, autoscaling also becomes more LLM-specific. HeteroScale studies autoscaling in heterogeneous and disaggregated LLM inference and proposes a coordinated policy to jointly maintain balance between prefill and decode pools \cite{R7}. TokenScale introduces token velocity as a leading indicator for proactive autoscaling and further proposes convertible decoders that can temporarily absorb prefill demand during bursts \cite{R8}. These studies suggest that autoscaling for LLMs should be co-designed with the serving architecture instead of being inherited directly from microservice-oriented scaling rules \cite{R7}\cite{R8}.

\subsubsection{Heterogeneous Resource-Aware Placement}
\label{sec:heterogeneous-placement}

Heterogeneous resource-aware placement asks which hardware is most suitable for each component of an LLM workload. Even within one deployment, long-context inference, adapter execution, expert routing, and lightweight preprocessing expose very different compute, memory, and communication characteristics. Modern serving systems therefore move beyond a flat accelerator-centric view of the cluster and instead map different workload components to different hardware roles \cite{R4}\cite{R9}\cite{R10}\cite{R11}\cite{R13}. 

Representative systems make this trend explicit. LoongServe targets long-context serving and uses elastic sequence parallelism to adapt the degree of parallelism across requests and phases \cite{R9}. dLoRA dynamically orchestrates requests and adapters for LoRA serving, including dynamic merge/unmerge and request-adapter co-migration across worker replicas \cite{R10}. Punica improves multi-tenant LoRA serving by sharing the base model and batching multiple LoRA adapters efficiently \cite{R13}. For MoE serving, MegaScale-Infer disaggregates attention and FFN modules, enabling independent scaling, tailored parallel strategies, and heterogeneous deployment \cite{R11}. Together, these systems show that heterogeneous placement for LLMs is not merely about choosing between CPUs and GPUs, but about exposing model-internal structure and mapping each component to the hardware class that best matches its bottleneck.

\subsubsection{Topology- and Communication-Aware Placement}
\label{sec:topology-aware}

Topology-aware placement begins from the observation that the performance of large LLM systems is often constrained more by tensor movement than by the peak arithmetic throughput of individual accelerators. Parallel serving strategies such as tensor parallelism, pipeline parallelism, MoE routing, and phase-disaggregated inference all depend on communication locality. AlpaServe demonstrates that model parallelism can be combined with statistical multiplexing to improve serving efficiency under bursty workloads, but its gains still depend on coordinated placement and communication \cite{R12}. Similarly, DistServe, Splitwise, and MegaScale-Infer all rely on communication-efficient placement because the benefit of decomposition diminishes when cross-node transfer becomes dominant \cite{distserve} \cite{R4}\cite{R11}. 

The implication for LLM serving is clear. Co-locating tightly coupled shards, adjacent stages, or frequently activated experts on nearby accelerators reduces synchronization overhead and improves both throughput and tail latency. In contrast, treating the cluster as a flat resource pool and scattering components arbitrarily increases communication cost and destabilizes elastic scheduling. As deployments continue to scale, topology-aware placement is becoming a prerequisite for predictable performance rather than a secondary optimization \cite{distserve} \cite{R4}\cite{R11}\cite{R12}.

\subsubsection{Load Balancing and Straggler Mitigation}
\label{sec:load-balancing}

At the request level, LLM workloads are shaped by skew and straggler behavior rather than by average-case regularity. Prompt length, generation length, tool invocation, and decoding settings vary substantially across requests, so naive request assignment often leaves some workers underutilized while others are blocked by a few long-running jobs. Existing serving systems therefore combine batching, queueing, routing, and preemption mechanisms to reduce imbalance \cite{orca} \cite{vllm} \cite{R6}\cite{R14}\cite{BanaServe}. 

Orca introduced iteration-level scheduling for autoregressive transformer serving \cite{orca} . vLLM improved shared serving throughput through PagedAttention-based KV-cache management \cite{vllm} . FastServe reduced head-of-line blocking by enabling preemption at the granularity of each output token \cite{R6}. Niyama extends this line of work by introducing fine-grained QoS classes, dynamic chunking, hybrid prioritization, and selective relegation for shared LLM serving \cite{R14}. Taken together, these studies suggest that load balancing and straggler mitigation should be treated as part of a unified multi-layer scheduling framework spanning request routing, phase placement, and cluster-level autoscaling. As summarized in Table \ref{tab:llm_serving}, these methods mainly differ in scheduling granularity, resource coupling, and optimization target.

\begin{table}[!ht]
    \centering
    \caption{Comparison of Key Scheduling Methods for Heterogeneous LLM Serving}
    \begin{tabular}{|c|c|c|c|}
    \hline
        \textbf{Method} & \textbf{Core idea} & \textbf{Advantage} & \textbf{Limitation } \\ \hline
        Monolithic serving & Unified prefill and decode pool & Simple architecture & Strong phase interference  \\ \hline
        P/D disaggregation & Separate prefill and decode & Balances TTFT and TPOT & Transfer and routing overhead  \\ \hline
        Heterogeneous placement & Match tasks to hardware & Higher utilization & Requires accurate profiling  \\ \hline
        Topology-aware placement & Place modules by interconnect & Lower communication cost & Sensitive to cluster topology  \\ \hline
        Request-level scheduling & Use batching and queueing & Mitigates stragglers & Hard under dynamic traffic  \\ \hline
    \end{tabular}
    \label{tab:llm_serving}
\end{table}

\subsection{Energy-Aware and Carbon-Aware Workload Placement}
\label{sec:energy-aware}

\color{blue}The energy- and carbon-aware mechanisms discussed in this section directly
respond to the energy-efficiency and sustainability challenges identified in
Sections~3.1 and~3.3. Rather than treating energy consumption as a secondary
objective, these approaches incorporate power, thermal, and carbon signals into
resource placement and scheduling decisions.\color{black}

Energy and carbon considerations have become central to LLM deployment because both large-scale training and globally distributed inference consume substantial electricity. Conventional scheduling policies that optimize only latency or monetary cost overlook an increasingly important factor: the relation between computation and the carbon intensity of the surrounding infrastructure. The challenge is to reduce energy consumption and carbon emissions without undermining application-level performance guarantees. 

This section examines four complementary aspects of energy-aware and carbon-aware management: energy-efficient hardware allocation, data-center and region selection, carbon footprint monitoring and budgeting, and integration into cloud-native control planes.

\subsubsection{Energy-Efficient Hardware Allocation}
\label{sec:energy-hardware}

At the hardware level, different accelerators provide very different performance per watt for LLM workloads. Newer generations may offer higher throughput and stronger memory bandwidth under similar or lower power budgets, but real efficiency depends strongly on execution configuration, including precision, kernel implementation, and batching policy. Splitwise is particularly relevant here because it shows that the compute-intensive prompt phase and the memory-intensive generation phase can be mapped to different machines with distinct cost and power characteristics \cite{R4}.  DynamoLLM and throttll’em~\cite{DynamoLLM}\cite{throttlelm} optimize the energy consumption of LLM service via throttling the GPU clock frequency and minimizing the resource usage on the fly.

\subsubsection{Data Center and Region Selection}
\label{sec:datacenter-selection}

Energy-aware scheduling extends beyond individual nodes to the level of data centers and regions. Different regions vary in power-source composition, cooling efficiency, and real-time grid carbon intensity. For non-urgent workloads such as pre-training, offline evaluation, or non-interactive fine-tuning, carbon-aware scheduling can shift computation to regions and time windows with lower carbon intensity. Even for latency-sensitive inference, regional balancing may still reduce emissions while respecting latency budgets. The key insight is that regional placement should be treated not only as a performance decision but also as a sustainability decision. 

\subsubsection{Carbon Footprint Monitoring and Budgeting}
\label{sec:carbon-monitoring}

Carbon-aware control depends on accurate telemetry. For LLM workloads, this means estimating how much energy and how many emissions are attributable to a training run, an inference deployment, or an individual tenant. This is difficult in shared environments where power is often measured at rack level or data-center level while multiple workloads run concurrently. Recent work on LLM power characterization shows that proper telemetry can expose significant oversubscription and scheduling opportunities. Microsoft’s study analyzes power behavior for several popular LLM configurations and proposes POLCA, showing that power-aware control can safely increase effective datacenter capacity \cite{R17}. 

\subsubsection{Integration into Cloud-Native Control Planes}
\label{sec:cloud-integration}

For energy-aware and carbon-aware strategies to have operational impact, they must be incorporated into the control planes that perform everyday scheduling decisions. In practice, this means extending cluster managers, orchestrators, and autoscalers so that power, temperature, and carbon-related signals become inputs to placement and scaling logic. TAPAS extends this line of work by jointly considering thermal and power constraints in cloud platforms, showing that LLM scheduling must increasingly account for cooling, thermal throttling, and emergency management in addition to latency and throughput \cite{TAPAS}. Sustainable LLM scheduling should therefore be understood as a multi-objective optimization problem rather than as an isolated environmental add-on. The main optimization levels are summarized in Table \ref{tab:energy_aware}.

\begin{table}[!ht]
    \centering
    \caption{Main Optimization Levels in Energy- and Carbon-Aware LLM Scheduling}
    \resizebox{1.0\textwidth}{!}{%
    \begin{tabular}{|c|c|c|c|}
    \hline
        \textbf{Level} & \textbf{Main objective} & \textbf{Typical approach} & \textbf{Main challenge } \\ \hline
        Hardware level & Improve performance per watt & Select hardware and precision & Workload-sensitive efficiency  \\ \hline
        Cluster level & Reduce power and thermal pressure & Power-aware placement and routing & Cross-node coordination  \\ \hline
        Region level & Reduce carbon emissions & Cross-region and time shifting & Latency constraints  \\ \hline
        Control-plane level & Integrate sustainability into scheduling & Use power and carbon signals & Complex multi-objective trade-offs  \\ \hline
    \end{tabular}
    }
    \label{tab:energy_aware}
\end{table}

\subsection{Adaptive QoS for Multi-Tenant Environments}
\label{sec:adaptive-qos}

\color{blue}The adaptive QoS mechanisms in this section address the multi-tenant resource
allocation, fairness, and isolation challenges identified in Section~3.3.
Because LLM workloads differ substantially in latency requirements, request
characteristics, and resource intensity, effective multi-tenant serving requires
resource differentiation and isolation mechanisms that are sensitive to
LLM-specific service behavior.\color{black}

Cloud-native LLM deployments are inherently multi-tenant. Multiple applications, teams, or customers share the same infrastructure, often through a common service platform. Such consolidation improves utilization, but it also makes quality of service more difficult to guarantee. LLM workloads intensify this problem because a small number of heavy tenants or unexpectedly popular features can saturate shared accelerators and degrade performance for others, especially when long contexts, retrieval, or tool usage are involved. 
\color{blue}This section examines four building blocks of adaptive QoS in multi-tenant LLM environments: priority- and SLA-aware allocation, request classification and differentiated handling, fairness and resource isolation, and feedback-driven adaptation.\color{black}

\subsubsection{Priority-Aware and SLA-Aware Resource Allocation}
\label{sec:priority-sla}

In multi-tenant LLM deployment, workloads differ substantially in business importance and service sensitivity. Some applications require strict tail-latency guarantees or strong availability, whereas others can tolerate slower service. Priority-aware and SLA-aware allocation mechanisms make these differences explicit by allowing tenants to specify performance objectives such as latency targets, throughput guarantees, and availability requirements. In adapter-centric deployment, Punica shows that shared-base-model serving can significantly improve memory efficiency and throughput, but such shared settings also make explicit resource differentiation increasingly necessary \cite{R13}.  AdaServe~\cite{AdaServe} constructs a speculation tree tailored to each request’s latency target, via introducing a speculate-select-verify pipeline that enables fine-grained control over decoding speed while maximizing system throughput. Laser~\cite{AdaServe} seamlessly integrates inter-instance request dispatching with layer-level scheduling within instances, delivering high serving throughput with SLO guarantees.

\subsubsection{Request Classification and Differentiated Handling}
\label{sec:request-classification}

Even within the same tenant, requests may have very different service requirements. Interactive chat, streaming generation, offline batch processing, evaluation workloads, and background maintenance tasks should not be treated as a single undifferentiated class. Niyama argues that siloed infrastructure with only coarse interactive-versus-batch segregation is insufficient for modern LLM workloads. It introduces fine-grained QoS classification, dynamic chunking, hybrid prioritization, and selective relegation to support more precise latency-aware co-scheduling on shared infrastructure \cite{R14}. 

\subsubsection{Fairness and Isolation}
\label{sec:fairness-isolation}

Priorities and service classes are not sufficient unless the system can also enforce fairness in the use of shared accelerators. Without protection, a small number of aggressive or misconfigured tenants may consume a disproportionate share of available resources and leave others persistently underserved. Fairness and isolation mechanisms address this risk through weighted queueing, quota enforcement, rate limiting, and per-tenant limits on concurrency, GPU time, or memory usage. In practice, strict isolation is often too costly because fixed partitioning can produce fragmentation and low utilization, so many systems adopt soft isolation while preserving starvation protection under load \cite{R13}\cite{R14}. Virtual Token Counter algorithm~\cite{Fairness} ensures equitable resource allocation among tenants in LLM inference, by focusing on token-level fairness. Specifically, it provides strict, provable fairness guarantees and prevents resource monopolization by individual users.


\subsubsection{Feedback-Driven Adaptation}
\label{sec:feedback-adaptation}

Static QoS configurations rarely remain effective under real workloads. Traffic patterns evolve, new models are introduced, and user behavior changes continuously. Feedback-driven adaptation closes the loop between observed performance and resource allocation. Controllers monitor latency, throughput, drop rate, tenant-level utilization, and other service indicators, then adjust concurrency limits, batching parameters, and capacity distribution when performance deviates from target levels. Niyama is representative here because it couples fine-grained QoS classes with real-time state-aware adaptation, allowing better capacity use under shared serving \cite{R14}. 
The broader implication is that QoS management for LLM systems must move beyond classical cloud abstractions. Prompt length, token streaming behavior, retrieval delay, and tool invocation all affect perceived service quality. Future QoS frameworks should therefore expose LLM-native service indicators and provide tenant-specific guarantees in forms that reflect the real behavior of generative workloads.

\subsection{Use of Reinforcement Learning and AI-Driven Orchestration}
\label{sec:rl-orchestration}

\color{blue}The AI-driven orchestration mechanisms discussed in this section are motivated
by two limitations identified earlier: the inability of reactive resource
management to anticipate rapidly changing LLM workloads (Section~3.3) and the
lack of cross-layer visibility required for effective scheduling and resource
allocation (Section~3.2). Learning-based control and observability-aware
orchestration provide a potential means of closing this gap by incorporating
workload predictions, fine-grained system telemetry, and feedback into resource
management decisions.\color{black}

The growing complexity of LLM workloads has led to increasing interest in reinforcement learning and other AI-driven approaches to resource management. Hand-tuned heuristics struggle to handle the large control space of modern LLM systems, which includes scaling choices, placement decisions, batching policies, precision settings, routing strategies, and heterogeneous hardware coordination. Learning-based controllers offer the possibility of adapting over time, capturing interactions among control variables, and optimizing several objectives simultaneously.

This section examines four roles that reinforcement learning and AI can play in LLM resource management: RL-based autoscaling and scheduling, AI-driven orchestration across the stack, predictive and proactive resource management, and observability-aware closed-loop control.

\subsubsection{RL-Based Autoscaling and Scheduling}
\label{sec:rl-autoscaling}

RL-based autoscaling and scheduling formulate resource management as a sequential decision problem. At each control step, the agent observes system state, including queue length, accelerator utilization, latency statistics, token generation rate, phase-specific load, and KV-cache pressure. It then chooses actions such as scaling a service up or down, relocating workload components, or modifying batching parameters. AWARE is a representative production-oriented framework in this direction. It provides an extensible RL-based autoscaling framework, leverages meta-learning and bootstrapping for safer adaptation, and shows that RL controllers can move closer to real deployment constraints~\cite{DRPC}. 

A further challenge lies in engineering integration. In real cloud-native systems, the controller does not act on a perfectly observable and instantaneous environment. Monitoring signals may be delayed, scaling actions require warm-up time, topology information may be incomplete, and communication bottlenecks may change faster than the control loop can react. As a result, RL schedulers must be designed with explicit awareness of observability delay, action delay, and infrastructure uncertainty. This is one reason why offline training, constrained control, and human oversight remain important in practical deployment \cite{R15}.

\subsubsection{AI-Driven Orchestration Across the Stack}
\label{sec:ai-orchestration}

Beyond low-level scaling and placement, AI models can orchestrate higher-level aspects of LLM services. This includes selecting which model or model variant to use for a given request, deciding when to apply compression or quantization, and routing requests to edge or central data centers. It also covers choices such as whether to serve a request from a cached result, a smaller distilled model, or a full-sized model with retrieval and tools.

AI-driven orchestration frameworks often treat these choices as contextual bandit or RL problems, where each request or workload carries context (features about the user, request, and system state), and the orchestrator learns which combination of actions yields the best trade-off between quality, latency, and cost. In multi-model or multi-LLM deployments, such orchestrators can significantly improve overall efficiency by avoiding over-provisioning high-end models for simple tasks and by steering heavy workloads to better-suited clusters or times of day.

\subsubsection{Predictive and Proactive Resource Management}
\label{sec:predictive-management}

Reactive policies that respond only to current load and utilization are often too slow for LLM workloads, where cold-start times can be long and traffic spikes can be abrupt. Predictive and proactive resource management adds forecasting to the loop. Time-series models can predict future request rates, token volumes, or mix of request types based on historical data and external signals (e.g., time of day, product launches). For LLMs, more specialized predictors can estimate the distribution of prompt lengths or the likelihood of tool-using requests, which directly affect resource usage.

These predictions can then drive proactive actions: spinning up additional replicas ahead of an expected surge, warming caches or KV-stores, pre-loading popular adapters or experts onto accelerators, or shifting non-urgent workloads out of high-demand windows. When combined with RL or other decision-making mechanisms, prediction becomes a way to reduce the need for aggressive exploration in production and to smooth the control problem the agent faces. TokenScale is representative of this trend because it uses token velocity as a leading signal for more timely autoscaling in disaggregated serving \cite{R8}. 

\begin{table}[!ht]
    \centering
    \caption{Summary of QoS and Intelligent Control Mechanisms in Multi-Tenant LLM Serving}
    \resizebox{1.0\textwidth}{!}{%
    \begin{tabular}{|c|c|c|c|}
    \hline
        \textbf{Mechanism} & \textbf{Main purpose} & \textbf{Advantage} & \textbf{Limitation } \\ \hline
        Priority- and SLA-aware allocation & Protect critical services & Clear service differentiation & Lower flexibility  \\ \hline
        Request classification & Separate request types & Better queueing and batching & Needs accurate classification  \\ \hline
        Fairness and isolation & Prevent resource monopolization & Higher system robustness & Strong isolation lowers utilization  \\ \hline
        Feedback-driven control & Adjust resources by QoS & Adapts to workload changes & Hard to ensure stability  \\ \hline
        RL/AI orchestration & Learn scaling and routing & Handles dynamic environments & Complex training and deployment \\ \hline
    \end{tabular}
    }
    \label{tab:QoS}
\end{table}

\subsubsection{Closing the Loop with Observability}
\label{sec:observability}

All learning-based resource management depends on strong observability. For LLM systems, observability includes fine-grained metrics on request latency, token throughput, accelerator utilization, memory use, KV-cache behavior, control actions, and policy outcomes. Such telemetry supports policy training, offline diagnosis, rollback for unsafe behavior, and explanation of controller decisions. In rapidly evolving LLM deployments, observability is therefore not merely an auxiliary monitoring capability. It is a foundational condition for safe and effective learning-based control \cite{R15}\cite{R16}. 

For LLM services, the most promising direction is not to replace all existing control logic with a fully learned system, but to build hybrid frameworks in which learning-based modules augment established scheduling mechanisms under explicit operational guardrails. The main QoS and intelligent control mechanisms are summarized in Table \ref{tab:QoS}.\\

\color{blue}\textbf{Takeaway.}
Existing techniques have significantly improved the scalability and efficiency of
LLM systems, including distributed execution, elastic resource management, and
serving optimization. However, most current approaches optimize individual system
components independently. Future research should explore cross-layer optimization
and intelligent orchestration mechanisms that integrate workload characteristics,
resource availability, and service objectives.\color{black}

\section{Emerging Trends and Innovations}
\label{ch:5}

\color{blue}This section explores the emerging trends and innovations in the field of LLMs that are enabled by advances in cloud-native architectures, distributed systems, and new computing paradigms.\color{black}

\subsection{Serverless Inference for LLMs}
\label{sec:serverless-inference}

Nowadays, serverless inference platforms, such as AWS Lambda or Google Cloud Functions, can enable LLMs to be served at scale with minimal operational overhead. This computing paradigm allows cloud tenants to focus on building their AI application, leaving the complex, undifferentiated heavy lifting of infrastructure management to the platform provider. In the context of LLMs, the principle of serverless LLM inference is to shift the operational burden, which are generated from resource scaling or fault tolerance, from the cloud users to the cloud provider. For example, the serverless platform automatically finds an available GPU server to execute the computation load for the incoming inference requests at peak time.

\color{blue}In this part, we investigate three perspectives of the serverless LLM inference. Section~\ref{sec:tco-oriented} illustrates how existing serverless LLM inference works reduce the total cost of ownership (i.e., TCO) using various methods, which becomes increasingly important due to the high energy consumption and expensive hardware cost of GPU devices. Section~\ref{sec:scalability} describes the considerations of service scalability. To be specific, it discusses how existing works leverage the advantages of serverless LLM inference to enhance the service provisioning to a large scale. Section~\ref{sec:simplified-management} covers the management simplification, where functionalities such as fault tolerance and programmability can be significantly improved when integrated with the serverless workflow.\color{black} 

\subsubsection{TCO-oriented service provision}
\label{sec:tco-oriented}

The total cost ownership is one of the most important considerations for cluster administrators that provision LLM services. This is due to the massive demands in computation and storage resources, which are naturally introduced by hosting the LLM in the cloud. Resource allocation is a promising direction towards low TCO. To be specific, advanced systems like DeepBAT ~\cite{DeepBAT} use deep learning techniques to explore the optimal serverless configurations (i.e., batch size among instances) to meet the service level objectives (SLOs) goals while keeping costs minimal. Hetis ~\cite{hetis} proposes efficient strategies to harness low-end resources, which helps reduce the operational cost in LLM inference service. ServerlessLLM ~\cite{ServerlessLLM} mitigates the cold start problem via incorporating a new checkpoint format design, which accelerates the parameter loading process and therefore prevents resource oversubscription.

On the other hand, the thermal and power management also benefits the TCO reduction in the cluster level. TAPAS ~\cite{R18} balances the thermal conditions and energy consumption across servers, thereby enhancing the cooling capability and the power supply in the cluster. DynamoLLM ~\cite{DynamoLLM} and Thrott'LLeM ~\cite{throttLLeM} reduce the energy consumption via dynamically adjusting the GPU clock frequency, parallelization configuration, and request dispatching among instances. To this end, TCO under these system is consistently reduced.

\begin{figure}
    \centering
    \includegraphics[width=0.8\linewidth]{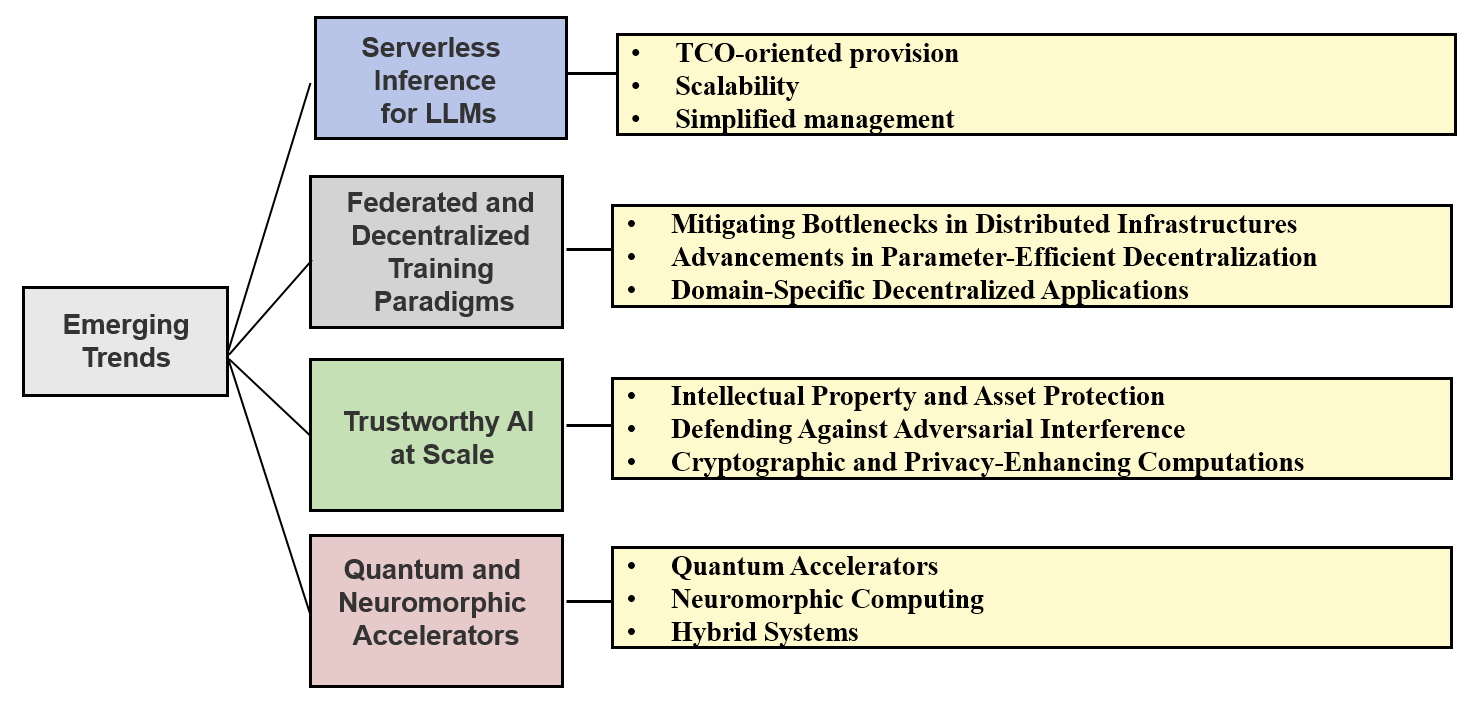}
    \caption{Main emerging trends in system support for LLMs}
    \label{fig:emerging}
\end{figure}

\subsubsection{Scalability}
\label{sec:scalability}

Scalability is also a critical issue in the inference service, particularly in a large-scale cluster. Due to the light-weighted operational overhead, the service provider can spend less effort to predict the future situation for successful scaling. In the context of LLM inference, this property is even more important, since both the request arrival events and context length of requests are highly random, making the accuracy is difficult to guarantee.

Moreover, Medusa ~\cite{Medusa} materializes the CUDA graphs as well as the information needed by the KV cache initialization in the offline phase, and restores them efficiently in the online phase. By this design, it mitigates the unique complexity caused by LLM inference, i.e., the capturing stage, which is responsible for dynamically constructing CUDA graphs for different batch sizes. Furthermore, BlitzScale ~\cite{blitzscale} leverages the network-based model-aware multicast techniques to speed up the model loading process, and introduces model-aware remote execution to live migrate the data plane. As such, it mitigates the cold start problem in serverless LLM inference, reducing tail latency and therefore enhancing the scalability at peak time.

\subsubsection{Simplified management}
\label{sec:simplified-management}

Another critical advantage of serverless LLM inference lies in the management simplification. For example, the serverless paradigm effectively decouples the service deployment from the faults from both software and hardware, making the inference service more robust. More specifically, when faults take place in one inference instance, requests in other instances will not be impacted under the serverless paradigm.

Moreover, programmability is also an important management issue, which can be facilitated by the clear workflow definition in serverless LLM inference. Pie ~\cite{pie} decomposes the token generation process of LLM inference into modular API handlers, allowing user-provided programs to conduct resource orchestration on the entire workflow. In this sense, it enables fine-grained control over KV cache management (e.g., modular scaling), custom decoding procedures, and tight integration of external computation and I/O workflow.

\subsection{Federated and Decentralized Training Paradigms}
\label{sec:federated-training}

The integration of massive neural architectures into decentralized ecosystems introduces a transformative approach to machine learning, primarily driven by the necessity to avoid centralizing sensitive information. Conventional centralized optimization methodologies require aggregating immense datasets onto a single server, which frequently violates strict legal frameworks and user confidentiality expectations. In contrast, decentralized paradigms empower participants to refine algorithms using their localized, private datasets. Throughout this collaborative process, only the computed gradient updates or structural adjustments are transmitted to a coordinating server, successfully isolating the raw information from external exposure. This architectural shift not only perfectly aligns with contemporary data protection mandates but also capitalizes on the diverse information silos distributed across various institutions without compromising security.

\subsubsection{Mitigating Bottlenecks in Distributed Infrastructures}

Deploying these massive architectures across decentralized networks presents substantial systemic obstacles, particularly regarding communication bandwidth and endpoint hardware capabilities. Transmitting the billions of parameters typical of modern models during iterative synchronization rounds incurs severe latency and easily saturates available network bandwidth. Furthermore, edge endpoints frequently suffer from a profound lack of memory capacity, which strictly prohibits standard full-parameter optimization procedures. Beyond these hardware constraints, the statistical variance of datasets across different participants creates significant algorithmic hurdles. This inherent heterogeneity often disrupts the synchronization process, leading to divergent model weights, prolonged convergence times, and a deteriorated generalization capacity of the aggregated global framework.

\subsubsection{Advancements in Parameter-Efficient Decentralization}

To circumvent the aforementioned infrastructural limitations, contemporary systems research has aggressively adopted parameter-efficient optimization strategies within decentralized topologies. Instead of updating the entire neural network, these methodologies freeze the primary architecture and solely adjust a minuscule fraction of task-specific variables. Techniques involving low-rank matrix approximations decompose massive weight updates into mathematically manageable dimensional spaces, dramatically reducing both the memory footprint and the volume of data that must be communicated over the network. Similarly, alternative strategies involving the algorithmic optimization of input embeddings or the insertion of compact, trainable modules between transformer layers ensure that devices with highly constrained computational power can meaningfully participate in the collaborative refinement process.

\subsubsection{Domain-Specific Decentralized Applications}

Decentralized optimization frameworks have demonstrated immense potential across various specialized industries where data confidentiality is an absolute prerequisite. In the healthcare sector, clinical facilities possess highly sensitive patient records that cannot be legally aggregated into a central repository. By leveraging localized refinement, medical institutions can collaboratively enhance diagnostic reasoning and clinical decision-support platforms while maintaining total patient anonymity. Similarly, financial organizations utilize these decentralized networks to collaboratively improve risk assessment and market analysis algorithms without exposing proprietary trading strategies or heavily regulated financial profiles. These practical implementations underscore the viability of decentralized paradigms in fostering industry-wide AI advancements while strictly adhering to data governance laws.

\subsection{Trustworthy AI at Scale}
\label{sec:trustworthy-ai}

As AI systems are deployed across expansive and often untrusted networks, guaranteeing their operational integrity and security becomes a critical priority. The distribution of pre-trained frameworks to numerous client nodes inherently magnifies the attack surface, creating profound vulnerabilities for both the infrastructure and the intellectual property embedded within the neural architectures. Addressing these multifaceted vulnerabilities demands a comprehensive security paradigm that spans the entire lifecycle of the system, from the initial distribution of the neural weights to the execution of localized computations, ensuring that the deployment remains robust against malicious exploitation while preserving user confidentiality.

\subsubsection{Intellectual Property and Asset Protection}

The development of advanced artificial intelligence requires astronomical financial investments and extensive computational resources, rendering the resulting models highly valuable proprietary assets. When these sophisticated frameworks are disseminated in a white-box manner to decentralized participants for localized refinement, malicious actors are presented with the direct opportunity to reverse-engineer, clone, or misappropriate the foundational architecture. To neutralize these severe threats to intellectual property, infrastructure engineers are increasingly researching advanced obfuscation protocols. Mechanisms such as embedding traceable digital watermarks within the generated outputs and utilizing encrypted transmission channels are being actively deployed to securely deliver model parameters. These protective layers allow legitimate clients to interact with and refine the model without gaining unauthorized access to its underlying structural secrets.

\subsubsection{Defending Against Adversarial Interference}

Beyond the theft of proprietary designs, the utilization of open-source frameworks in decentralized computing environments introduces critical operational hazards, especially when end-users lack sophisticated cybersecurity expertise. Inadequately secured client nodes can be easily compromised, serving as entry points for threat actors to inject malicious data during the collaborative training phases. Such adversarial poisoning can seamlessly introduce latent vulnerabilities or backdoors into the globally aggregated model, ultimately compromising the reliability of the entire network. Consequently, neutralizing these risks necessitates the implementation of rigorous validation protocols capable of identifying and isolating anomalous gradient updates before they corrupt the central system, thereby safeguarding the collective intelligence from coordinated sabotage.

\subsubsection{Cryptographic and Privacy-Enhancing Computations}

Achieving genuine trustworthiness requires embedding privacy-preserving mathematical guarantees directly into the foundational computational workflows. To protect the granular details of individual datasets from being reconstructed through reverse gradient analysis, researchers are incorporating sophisticated cryptographic techniques, such as localized noise injection, during the optimization phase. While these mathematical perturbations successfully obscure sensitive records, they frequently introduce significant computational friction and can degrade the ultimate predictive precision of the algorithm. Furthermore, the adoption of isolated hardware execution environments is becoming vital for processing highly sensitive inferences at the network edge. By combining these trusted execution zones with rigorous access controls and auditable operational logs, distributed platforms can effectively shield both the localized training data and the proprietary model logic from unauthorized interception.

\subsection{Quantum and Neuromorphic Accelerators in Cloud-Native Settings}
\label{sec:quantum-neuromorphic}

The continuous scaling of LLMs has begun to expose the fundamental efficiency limits of classical computing infrastructures, particularly in terms of energy consumption, parallelism granularity, and optimization complexity. Emerging computing paradigms, including quantum and neuromorphic systems, offer a transformative pathway toward overcoming these constraints. When integrated into cloud-native distributed environments, these accelerators introduce novel computational abstractions that fundamentally reshape how large-scale models are trained and deployed. Rather than serving as isolated hardware innovations, they represent a shift toward heterogeneous, hybrid infrastructures capable of supporting the next generation of intelligent systems.

\subsubsection{Quantum Acceleration for Large-Scale Optimization}

Quantum computing introduces a fundamentally different computational paradigm based on superposition and entanglement, enabling the exploration of high-dimensional solution spaces with unprecedented efficiency. In the context of LLMs, quantum accelerators hold significant promise for addressing the computational bottlenecks inherent in large-scale optimization and probabilistic inference. Certain classes of optimization problems, such as parameter search in highly non-convex landscapes or sampling from complex distributions, can theoretically benefit from exponential or polynomial speedups when mapped to quantum algorithms.

However, integrating quantum processors into distributed cloud environments presents substantial challenges. Current quantum hardware remains constrained by limited qubit counts, noise sensitivity, and restricted coherence times, necessitating hybrid execution models where quantum circuits are tightly coupled with classical control systems. This introduces new orchestration complexities, as cloud-native schedulers must coordinate heterogeneous workloads spanning classical GPUs and remote quantum processing units, while minimizing communication latency and ensuring efficient task partitioning.

\subsubsection{Neuromorphic Computing for Energy-Efficient Inference}

In parallel, neuromorphic computing offers an alternative paradigm inspired by the structural and functional principles of biological neural systems. By leveraging event-driven computation and spiking neural representations, neuromorphic processors achieve significantly higher energy efficiency compared to conventional von Neumann architectures. This characteristic makes them particularly well-suited for latency-sensitive and energy-constrained inference scenarios in large-scale LLM deployments.

Within cloud-native settings, neuromorphic accelerators can be utilized to offload specific components of LLM inference pipelines, such as attention approximations, sparse activations, or continual learning modules. Their inherently asynchronous execution model also aligns well with distributed, loosely coupled system architectures. Nevertheless, significant challenges remain in bridging the abstraction gap between transformer-based models and spiking neural representations, requiring advances in model conversion techniques, programming frameworks, and runtime support.

\subsubsection{Hybrid Heterogeneous Architectures}

Recognizing the complementary strengths of classical, quantum, and neuromorphic systems, recent research has increasingly focused on hybrid architectures that integrate multiple computational paradigms within a unified cloud-native framework. In such systems, classical accelerators handle deterministic and large-scale tensor computations, quantum processors are selectively invoked for complex optimization or sampling tasks, and neuromorphic units provide energy-efficient execution for sparse or event-driven workloads.

This hybridization introduces new system-level design challenges, particularly in terms of workload decomposition, cross-platform scheduling, and consistency management. Efficiently partitioning LLM pipelines across heterogeneous resources requires fine-grained understanding of computational characteristics and data dependencies. Moreover, cloud-native orchestration layers must evolve to support dynamic resource discovery, adaptive execution planning, and cross-layer optimization across fundamentally different computing substrates.

\subsubsection{Emerging Applications and System Implications}

The integration of quantum and neuromorphic accelerators into cloud-native infrastructures opens up new opportunities for advancing LLM capabilities across diverse application domains. In scientific computing, hybrid quantum-classical workflows can accelerate complex simulations and knowledge discovery processes. In edge-cloud scenarios, neuromorphic processors enable ultra-low-power intelligent services, supporting real-time language understanding and decision-making in resource-constrained environments. 

These emerging applications highlight a broader system implication: future LLM infrastructures will increasingly rely on deeply heterogeneous, multi-paradigm computing environments. Realizing this vision requires not only advances in hardware technologies but also significant innovations in distributed systems design, including unified programming abstractions, interoperability standards, and intelligent orchestration mechanisms that can fully exploit the capabilities of next-generation accelerators.\\


\color{blue}\textbf{Takeaway.}
Sections~5.1--5.4 illustrate that the next stage of LLM infrastructure evolution is being shaped by several complementary trends. Serverless inference seeks to improve elasticity, scalability, cost efficiency, and operational simplicity; federated and decentralized training extends LLM development across distributed and privacy-sensitive data sources; trustworthy AI mechanisms address the security, integrity, privacy, and intellectual-property risks introduced by increasingly distributed execution; and quantum, neuromorphic, and hybrid heterogeneous computing explore longer-term opportunities for improving computational and energy efficiency. Collectively, these trends suggest a transition toward LLM infrastructures that are more elastic, decentralized, trustworthy, and heterogeneous. At the same time, they introduce new systems challenges in state management, communication efficiency, security enforcement, resource discovery, interoperability, and cross-platform orchestration, which motivate the research agenda developed in the following section.\color{black}

\color{blue}
\subsection{Cross-sectional Synthesis of Challenges, Strategies, and Emerging Trends}

\begin{table*}[t]
	\centering
	\caption{\color{blue}Transversal synthesis linking major LLM system challenges, current
		strategies, remaining limitations, and emerging research trends.}
	\label{tab:cross_section_synthesis}
	{\color{blue}
	\renewcommand{\arraystretch}{1.12}
	\footnotesize
	\begin{tabularx}{\textwidth}{
			p{2.65cm}
			p{3.65cm}
			p{4.05cm}
			X}
		\toprule
		\textbf{Challenge from Section 3} &
		\textbf{Current Techniques/Strategies from Section 4} &
		\textbf{Remaining Limitation/Gap} &
		\textbf{Emerging Trends from Section 5} \\
		\midrule
		
		Hardware heterogeneity and communication overhead
		&
		Heterogeneous and topology-aware placement; elastic scheduling; P/D disaggregation
		&
		Limited cross-layer adaptation; topology and workload dynamics remain difficult
		to optimize jointly
		&
		Hybrid heterogeneous infrastructure; emerging accelerators
		\\
		
		Reactive resource management and bursty workloads
		&
		Autoscaling; predictive and proactive provisioning; RL-based scheduling
		&
		Prediction uncertainty, control delay, and deployment complexity
		&
		Serverless inference; AI-driven orchestration
		\\
		
		High energy consumption and TCO
		&
		Energy- and carbon-aware placement; power-aware scheduling
		&
		Joint optimization of latency, cost, energy, and carbon remains challenging
		&
		Serverless inference; energy-efficient emerging hardware
		\\
		
		Multi-tenant QoS, fairness, and isolation
		&
		SLA-aware allocation; request classification; fairness and isolation;
		feedback-driven control
		&
		Strong isolation may reduce utilization; diverse LLM workloads complicate
		SLO enforcement
		&
		Decentralized and federated services; finer-grained intelligent orchestration
		\\
		
		Cross-layer visibility and observability
		&
		Fine-grained telemetry; observability-aware control
		&
		Incomplete or delayed signals and difficulty converting telemetry into reliable
		control decisions
		&
		AI-driven orchestration; autonomous system management
		\\
		
		Privacy and data governance
		&
		Secure execution; access control; privacy-preserving methods
		&
		Privacy mechanisms may introduce computational and communication overhead;
		end-to-end governance remains difficult
		&
		Federated and decentralized training
		\\
		
		Distributed data movement and locality
		&
		Caching; data-aware placement; retrieval optimization
		&
		Data movement, consistency, and synchronization remain costly at scale
		&
		Federated and decentralized intelligence; edge-cloud systems
		\\
		
		Limited computational efficiency
		&
		Parallelism; quantization; sparsity; specialized acceleration
		&
		Hardware efficiency and model--system co-design remain constrained
		&
		Quantum and neuromorphic computing; advanced heterogeneous accelerators
		\\
		
		Fragmented interfaces and deployment models
		&
		Model formats; APIs; deployment abstractions
		&
		Cross-layer interoperability and evolving standards remain incomplete
		&
		Heterogeneous and interoperable cloud-native AI ecosystems
		\\
		
		\bottomrule
	\end{tabularx}
}
\end{table*}

Sections 3--5 examine the evolution of cloud-native LLM systems from complementary
perspectives. Section~3 identifies the fundamental challenges introduced by LLM
workloads, Section~4 reviews representative techniques and strategies that address
these challenges, and Section~5 discusses emerging approaches that may extend the
capabilities of current infrastructures. To make the relationships among these
perspectives explicit, Table~\ref{tab:cross_section_synthesis} provides a
transversal synthesis linking the major challenges with representative current
techniques, their remaining limitations or gaps, and emerging trends that may
contribute to addressing those gaps.

As summarized in Table~\ref{tab:cross_section_synthesis}, the relationship between
a challenge and an emerging trend is not necessarily one-to-one. Some emerging
approaches can strengthen existing solutions, whereas others address a particular
limitation while introducing additional system-level requirements. For example,
serverless inference can simplify resource provisioning and improve elasticity
for bursty workloads, but cold-start overhead, model-state management, and
accelerator initialization remain important challenges. These limitations
motivate further research on state-aware serverless runtimes and more efficient
resource provisioning mechanisms.

Similarly, federated and decentralized training can reduce the need to centralize
sensitive data, directly addressing privacy and data-governance concerns.
However, decentralization introduces new bottlenecks related to communication
bandwidth, endpoint memory capacity, data heterogeneity, and synchronization.
Parameter-efficient adaptation and communication-aware distributed execution
therefore become important complementary directions.

Emerging quantum and neuromorphic computing paradigms provide another example of
this dual role. They may offer new opportunities for computational efficiency or
energy-efficient execution, but their integration with cloud-native LLM
infrastructures introduces additional challenges in workload decomposition,
heterogeneous resource scheduling, programming abstractions, interoperability,
and orchestration. These directions should therefore be viewed as longer-term
opportunities for heterogeneous AI infrastructures rather than direct replacements
for current GPU-based systems.

Overall, Table~\ref{tab:cross_section_synthesis} reveals a broader transition from
isolated optimization toward coordinated, cross-layer system design. Existing
techniques increasingly address individual bottlenecks, while emerging trends
create opportunities to integrate computation, resource management, data,
security, and deployment decisions within unified control frameworks. More
importantly, the remaining limitations and gaps identified in the table provide
the basis for prioritizing future research directions. These gaps motivate the
research agenda presented in Section~6, where we discuss how cloud-native LLM
infrastructures can evolve toward more adaptive, intelligent, efficient, and
sustainable systems.

\color{black}

\section{RESEARCH AGENDA FOR CLOUD-NATIVE AND DISTRIBUTED LLM INFRASTRUCTURES}
\label{ch:6}

\color{blue}
The preceding sections establish a progression from the fundamental challenges of supporting LLM workloads to current system-level strategies and emerging infrastructure paradigms. Section~3 identifies the major computational, system and software, operational, privacy and security, data-management, and standardization challenges of cloud-native LLM systems. Section~4 examines representative resource-management and optimization mechanisms for addressing these challenges, while Section~5 discusses emerging paradigms that may further extend the capabilities of current infrastructures. The transversal synthesis at the end of Section~5 further reveals that important gaps remain between the capabilities of current techniques and the requirements of future large-scale LLM services.

\begin{figure}
    \centering
    \includegraphics[width=0.95\linewidth]{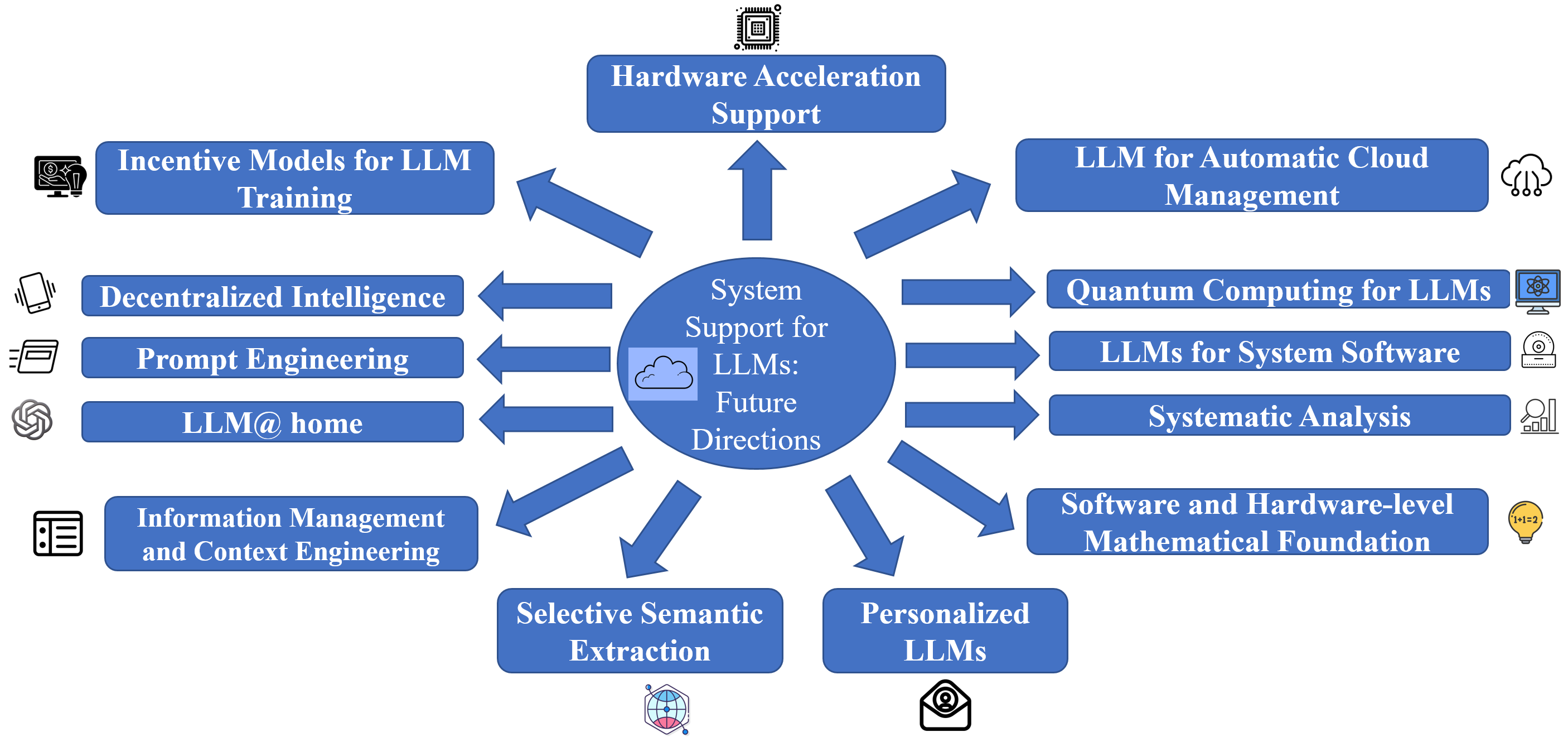}
    \caption{Research Agenda for Cloud-Native and Distributed System Support of LLMs}
    \label{fig:future}
\end{figure}

Building on these remaining gaps, this section develops a research agenda for the evolution of cloud-native and distributed LLM infrastructures as shown in~Fig. \ref{fig:future}. Rather than viewing the following topics as independent technological trends, we examine each direction from a common systems perspective: \textit{which previously identified limitation motivates the research direction, which infrastructure capability needs to evolve, which open research questions remain, and what system-level benefits may result}. In this sense, the research agenda is organized around the evolution of infrastructure capabilities rather than around individual technologies alone.

The resulting agenda spans multiple layers of the LLM infrastructure stack. At the computing layer, future systems require more efficient sparse computation, tighter software--hardware co-design, and mechanisms for incorporating increasingly heterogeneous accelerators. At the system-software and orchestration layers, LLMs themselves may become active components of scheduling, resource management, and autonomous cloud operation. At the data and state-management layers, prompts, retrieved information, KV caches, and persistent contextual memories increasingly need to be treated as first-class distributed resources. At the deployment layer, decentralized, edge-cloud, and personalized LLM services introduce new requirements for resource coordination, incentives, isolation, privacy, and lifecycle management. Sections~6.1--6.13 develop these directions in detail while maintaining this common infrastructure-oriented perspective.

Table~\ref{tab:research_agenda} summarizes the research agenda by linking each direction to its motivating gap, the infrastructure capability that needs to evolve, representative open research questions, and expected system-level benefits. This organization makes explicit how the research directions emerge from the challenges and remaining limitations identified in Sections~3--5. These emerging technologies and methodologies will shape the evolution of LLMs, enhancing their scalability, efficiency, and application across various domains. \color{black}


\begin{table*}[t]
	\centering
	\caption{\color{blue}Research agenda for cloud-native and distributed LLM infrastructures.}
	\label{tab:research_agenda}
	{\color{blue}
	\footnotesize
	\begin{tabularx}{\textwidth}{p{2.6cm}p{3.0cm}p{3.2cm}X}
		\toprule
		\textbf{Research Direction} &
		\textbf{Motivating Gap} &
		\textbf{Infrastructure Evolution} &
		\textbf{Representative Open Research Questions and Expected Benefits} \\
		\midrule
		
		Sparse and efficient computation
		&
		Memory- and computation-intensive LLM execution
		&
		Hardware-aware sparse execution and accelerator-aware runtimes
		&
		How can sparsity, memory layout, scheduling, and accelerator characteristics be jointly optimized? This may improve throughput, energy efficiency, and accelerator utilization. \\
		
		LLMs for system software
		&
		Static or narrowly optimized system-management policies
		&
		Reasoning-enhanced scheduling and system control
		&
		How can LLM reasoning be safely integrated with conventional optimization and control mechanisms to improve adaptability and resource efficiency? \\
		
		Software- and Hardware-Level Mathematical Foundations
		&
		Isolated optimization across algorithms, numerical formats, runtimes, and hardware
		&
		Cross-layer co-design and portable execution abstractions
		&
		How can computation, memory, communication, precision, and hardware characteristics be optimized jointly? \\
		
		Systematic analysis
		&
		Lack of comparable system-level understanding across workloads and platforms
		&
		Standardized measurement, benchmarking, and lifecycle evaluation
		&
		Which metrics and methodologies can reliably quantify performance, energy, cost, and sustainability across heterogeneous LLM deployments? \\
		
		Prompt engineering
		&
		Prompt growth increases prefill computation, KV-cache pressure, latency, and energy consumption
		&
		Prompt-aware orchestration and resource management
		&
		How should prompt compression and optimization react to runtime telemetry, memory pressure, latency targets, and energy budgets? \\
		
		Context Engineering
		&
		Context and memory state are expensive to recompute, transfer, and retain
		&
		Model-agnostic context representations, distributed caching, migration, and lifecycle management
		&
		How should contextual state be represented, placed, migrated, compressed, shared, and deleted across models and cloud-edge environments? \\
		
		LLM for Automatic Cloud Management
		&
		Reactive orchestration and increasing operational complexity
		&
		LLM-driven, predictive, verifiable, and autonomous control planes
		&
		How can autonomous management remain safe, explainable, SLO-aware, and auditable while reducing operational overhead? \\
		
		Decentralized Intelligence
		&
		Centralized infrastructures face cost, privacy, resilience, and geographic constraints
		&
		Distributed execution across heterogeneous and potentially untrusted nodes
		&
		How should decentralized schedulers handle communication, heterogeneity, verification, reliability, and resource incentives? \\
		
		Incentive Models for LLM Training
		&
		Distributed data and computing resources lack scalable contribution and reward mechanisms
		&
		Resource-aware valuation and incentive mechanisms integrated into distributed platforms
		&
		How can contributions be valued and rewarded efficiently while remaining robust to adversarial behavior, privacy constraints, and data withdrawal? \\
		
		Selective semantic extraction
		&
		Using large models for all structured-processing tasks can be costly and inefficient
		&
		Hierarchical LLM/SLM orchestration and distributed extraction pipelines
		&
		How should workloads be dynamically routed among heterogeneous models while preserving semantic fidelity, structured outputs, and security? \\
		
		LLM@Home
		&
		Cloud-only inference creates latency, privacy, availability, and bandwidth limitations
		&
		Edge-cloud and multi-device collaborative LLM infrastructure
		&
		How can heterogeneous home and edge devices jointly share computation and context while preserving privacy, access control, and energy efficiency? \\
		
		Personalized LLMs
		&
		Per-user state, adapters, and workloads create memory, fairness, isolation, and privacy challenges
		&
		Adapter-aware multi-tenant serving and personalized-state lifecycle management
		&
		How can millions of personalized states be scheduled, isolated, updated, and removed efficiently while maintaining QoS and privacy? \\
		
		 Quantum/Neuromorphic
		&
		Limits of conventional accelerators for selected optimization and sampling workloads
		&
		Heterogeneous runtimes capable of coordinating classical and emerging accelerators
		&
		Which LLM subroutines can benefit from emerging accelerators, and how should hybrid workloads be partitioned and scheduled? \\
		\bottomrule
	\end{tabularx}
}
\end{table*}

\subsection{Hardware Acceleration Support for Sparse Matrix Multiplications}
\label{sec:sparse-matrix}


\color{blue}The computational inefficiency and hardware heterogeneity identified in Sections~3.1 and~3.3 provide the primary motivation for this research direction. Existing cloud-native LLM infrastructures increasingly depend on heterogeneous accelerators, but sparse computation is still frequently optimized at the level of individual kernels or devices. The resulting gap is therefore not simply how to accelerate sparse matrix multiplication, but how distributed LLM infrastructures can expose and exploit sparsity while jointly considering accelerator characteristics, memory hierarchy, data movement, and workload placement.\color{black}

LLMs rely heavily on matrix operations, many of which remain memory-bound and computationally expensive on general-purpose hardware. 
SpMM has therefore become an important direction for reducing computational cost by exploiting zero-valued structures and improving effective arithmetic intensity. 
Recent studies show that efficient SpMM depends not only on sparsity itself, but also on architecture-aware designs that optimize memory access, dataflow organization, and hardware utilization. 
On GPUs, recent work has explored Tensor Core--adapted SpMM kernels that transform sparse operations into hardware-friendly blocked computations without requiring expensive matrix reordering~\cite{han2024tca_spmm}. 
On the reconfigurable-hardware side, FPGA-based designs and recent surveys demonstrate that customized dataflows, structured sparsity support, and memory-centric optimizations are critical for achieving high performance and energy efficiency in sparse linear algebra~\cite{liu2024fpga_spmm_survey}. 
More broadly, recent surveys on hardware accelerators for LLMs suggest that architecture-aware acceleration strategies will remain essential as LLM systems increasingly depend on heterogeneous hardware~\cite{kachris2025llm_hardware_survey}. 
Therefore, future progress in sparse LLM computation is likely to come from co-optimizing sparsity patterns, storage formats, and accelerator architectures rather than relying solely on software-level pruning.

\color{blue}From an infrastructure perspective, future systems therefore need to move from isolated sparse-kernel optimization toward sparsity-aware distributed execution. Important open questions include how cloud-native runtimes should describe accelerator-specific sparsity capabilities; how sparse and dense LLM operators should be dynamically mapped across heterogeneous GPUs, NPUs, and reconfigurable accelerators; how sparsity patterns, storage formats, memory traffic, and communication topology can be co-optimized; and how schedulers should react when model sparsity and hardware availability change dynamically. Addressing these questions could improve accelerator utilization, reduce memory and communication overhead, and provide more predictable performance and energy efficiency for large-scale LLM services.\color{black}



\subsection{LLMs for System Software}
\label{sec:llm-system-software}
\color{blue}
Existing resource managers mainly rely on predefined objectives, manually engineered heuristics, or specialized learning policies, whereas future LLM infrastructures will need to reason over heterogeneous telemetry, policies, workload semantics, and rapidly changing system states.\color{black}

LLMs are increasingly used to enhance cloud resource management, task scheduling, and system optimization by reasoning about dynamic workloads and guiding intelligent decision policies. For instance, active inference--based methods improve offloading and resource allocation in cloud--edge computing, outperforming traditional DRL approaches in adaptability and efficiency~\cite{he2024large}. Lightweight LLMs can assist reinforcement learning agents to generate superior task scheduling strategies in multi-cloud environments, reducing makespan and energy cost~\cite{tang2025llm}. Hybrid scheduling frameworks that combine LLM reasoning with learning agents exhibit better generalization across heterogeneous cloud conditions~\cite{pei2025llm}. Beyond cloud, green scheduling frameworks optimize model and data reuse across geographic data centers to cut emissions and costs~\cite{liu2025green}. Dynamic inference scheduling across distributed systems improves latency and resource balance~\cite{liang2026reliable}. Research on LLM-guided scheduling heuristics has shown improved resource utilization in fog environments~\cite{krishnamurthy2025scalable}\cite{srirama2026fog}. Taken together, these studies suggest that LLMs can evolve into higher-level reasoning components for system software, enabling more adaptive, context-aware, and efficient cloud-native infrastructures.

\color{blue} The research agenda is therefore to investigate how LLM reasoning can become a controlled component of system software rather than an external advisory tool. Open questions include how LLMs should interact with conventional optimizers, reinforcement-learning policies, and deterministic schedulers; which system states and telemetry should be exposed to LLM-based controllers; how reasoning latency and inference cost should be incorporated into scheduling decisions; and how incorrect or hallucinated recommendations should be detected before affecting production infrastructure. Resolving these issues could enable more adaptive system software capable of reasoning about workload semantics and operational context while preserving the predictability required by cloud-native services. \color{black}

\subsection{Software- and Hardware-Level Mathematical Foundations}
\label{sec:math-foundations}

\color{blue}Existing optimizations often improve one layer while shifting overhead to another. \color{black}The efficiency of LLMs depends not only on architectural innovations, but also on advances in algorithm design, numerical representation, and hardware-aware optimization. 
On the software side, recent surveys highlight the importance of scalable distributed training, efficient optimizer design, and memory-aware parallelization strategies for large models~\cite{wan2023efficient, zeng2025distributed}. 
On the numerical side, low-precision computation and quantization have become central techniques for improving throughput while maintaining model quality, especially in large-scale inference systems~\cite{zhao2024atom}. 
Recent systems such as Ladder further demonstrate that hardware-aware tensor transformations can systematically support efficient low-bit computation across modern accelerators~\cite{wang2024ladder}. 
More broadly, recent work on software--hardware co-design suggests that future machine learning systems should be optimized jointly across algorithms, data movement, runtime scheduling, and accelerator architectures to achieve both high performance and energy efficiency~\cite{vahdatpour2026codesign}. 
Together, these studies indicate that future LLM systems will rely on tightly integrated algorithmic, numerical, and architectural abstractions rather than isolated optimizations at a single layer.

\color{blue}A central research agenda is therefore to develop cross-layer abstractions through which model algorithms, numerical representations, runtime systems, and hardware can exchange optimization-relevant information. Open questions include how low-precision execution should interact with parallelism and scheduling, how memory and communication costs should be incorporated into algorithmic optimization, how portable abstractions can retain hardware-specific performance opportunities, and how system designers can jointly optimize accuracy, latency, throughput, memory, energy, and cost. Such foundations could move LLM infrastructure from collections of independently optimized layers toward coordinated end-to-end system co-design.\color{black}

\subsection{Systematic Analysis}
\label{sec:systematic-analysis}

The computational and operational requirements of LLMs have become a central research focus due to their extensive resource demands during training and inference. Systematic reviews highlight that electricity consumption and hardware demands scale substantially with model size, necessitating frameworks to estimate lifecycle energy use and sustainability challenges across training and inference stages~\cite{zhou2024survey}. Efficient inference strategies are categorized into data-, model-, and system-level optimizations that address memory and computational bottlenecks in resource-constrained environments~\cite{ji2026systematic}. Fernandes et al.~\cite{fernandez2025energy} analyse how inference efficiency optimizations affect real-world energy use across diverse hardware, emphasizing workload sensitivity to software stacks and accelerators. Narsepalle~\cite{narsepalle975energy} proposes energy-efficient training and adaptive inference techniques, demonstrating measurable reductions in computational workload and power consumption. Additionally, the work~\cite{wang2025ofq} on quantization and low-bit acceleration shows hardware approaches can boost efficiency without compromising performance. Broader systematic reviews also synthesize electricity demand challenges and multi-tier solutions for LLM operation~\cite{zhou2024survey}. Such analysis will help inform the future development of more efficient models and lead to better deployment strategies. A clear framework for understanding and optimizing LLMs will be key to scaling them effectively.

\color{blue}From the perspective of cloud-native and distributed infrastructure, the major unresolved issue is the lack of systematic and comparable methodologies for characterizing LLM system behavior across models, workloads, hardware platforms, deployment environments, and optimization techniques. Performance results obtained for one combination of model, accelerator, batch size, prompt distribution, and software stack may not generalize to another, while improvements in latency or throughput can introduce hidden costs in memory consumption, energy use, resource fragmentation, or operational complexity.

Future research should therefore establish systematic infrastructure-level evaluation methodologies for LLM systems. Open questions include which workload characteristics should be included in representative benchmarks; how prefill and decode behavior, KV-cache demand, communication traffic, energy, carbon emissions, and cost should be measured consistently; how optimization techniques should be compared across heterogeneous hardware; and how lifecycle costs spanning training, adaptation, deployment, and inference should be quantified. Standardized measurement methodologies and reproducible evaluation environments would provide an empirical foundation for comparing alternative infrastructures and would help transform isolated optimization results into generalizable system-design knowledge.\color{black}

\subsection{Prompt Engineering}
\label{sec:prompt-engineering}

\color{blue}Prompt engineering is traditionally considered an application- or model-level technique, but its growing influence on prefill computation, KV-cache consumption, communication volume, and end-to-end latency makes it increasingly relevant to cloud-native infrastructure. This direction is motivated by the resource-management and data-movement challenges identified in Sections~3 and~4: as prompts and system instructions become longer and more complex, their structure directly affects the amount of computation and memory required to serve each request. Prompt optimization should therefore be considered not only as a mechanism for improving response quality, but also as an infrastructure-level control variable.\color{black}

A vital future direction in this domain is automated prompt compression, which addresses the bandwidth and processing constraints of long-context workloads without requiring expensive model retraining. Recent advancements have introduced robust frameworks that systematically prune redundant or low-information tokens before they reach the resource-heavy primary LLM. For example, the ProCut framework utilizes attribution estimation to dynamically evaluate token importance, successfully compressing bloated prompts while maintaining, and sometimes enhancing, downstream task accuracy by preventing instruction dilution~\cite{xu2025procut}. Similarly, structured approaches leverage linguistic parse trees to guide the selective compression of prompts, adjusting the retention of tokens based on local information entropy and global hierarchical structures~\cite{mao2025parse}. This ensures that the core semantic meaning is preserved even at extreme compression ratios.

\color{blue}This perspective leads to a broader research agenda for infrastructure-aware prompt optimization. How should cloud-native runtimes expose the resource cost of prompts to application-level optimization mechanisms? How can prompt compression react dynamically to KV-cache pressure, queueing delay, accelerator availability, energy budgets, and SLOs? How should systems balance semantic fidelity against reductions in token processing and memory consumption? Can prompt representations and reusable prompt fragments be cached, versioned, and shared safely across requests or tenants? Addressing these questions would make prompt engineering part of the resource-management loop, allowing future LLM services to jointly optimize semantic quality and infrastructure efficiency.\color{black}

\subsection{Information Management and Context Engineering}
\label{sec:context-engineering}

\color{blue}Future LLM infrastructures should treat context as a first-class distributed system resource\color{black}. Context engineering focuses on enhancing the model's ability to retain and utilize information over time. Optimizing how LLMs manage large contexts and their memory during inference will improve their accuracy and response generation in real-world applications.

As LLMs increasingly power complex enterprise workflows, context engineering has formalized into a rigorous system-level discipline that treats the model's context window as a dynamic and critically evaluated resource. Traditional RAG pipelines often flood the context window with untargeted or irrelevant chunks, which can induce hallucinated outputs and degrade reasoning. To overcome this, cloud-native orchestration systems are shifting toward the intelligent evaluation and ranking of retrieved information before it is ever processed by the generative model. For instance, recent research introduces the concept of  Sufficient Context,  which formalizes a system's ability to automatically classify whether a retrieved document actually contains the necessary information to answer a query~\cite{joren2024sufficient}. By embedding selective generation directly into the orchestration layer, frameworks prevent LLMs from hallucinating when the context is inadequate. Furthermore, advanced systems like RankRAG demonstrate that instruction-tuning a single LLM to simultaneously perform context ranking and answer generation significantly outperforms traditional, decoupled retriever-reader architectures, ensuring only the highest-quality information occupies the strict memory budget~\cite{yu2024rankrag}.

Beyond evaluating static retrieval, optimizing information management requires dynamic orchestration between external data fetching and the model's internal long-term memory. Empirical studies comparing RAG against native long-context models reveal that while long-context models offer superior average performance for complex reasoning, their computational costs are often prohibitive~\cite{li2024retrieval}. Consequently, state-of-the-art orchestration now employs dynamic hybrid routing, which autonomously directs queries to either a lightweight RAG pipeline or a heavy long-context model based on real-time task complexity. For persistent, long-horizon applications, this routing is further augmented by dedicated  Memory Operating Systems.  Frameworks such as A-MEM (Agentic Memory) introduce self-evolving memory networks that autonomously extract, tag, and consolidate semantic knowledge over time~\cite{xu2025mem}.

\color{blue}The resulting research agenda is to treat context as a first-class distributed system resource rather than transient application input. Key open questions include how contextual state should be represented independently of a particular model; how prompts, retrieved documents, semantic memories, and KV states should be cached and placed across distributed infrastructure; when contextual state should migrate with a request across models or cloud-edge environments; how consistency and freshness should be maintained; and when contextual information should be compressed, archived, or deleted. Solving these problems could substantially reduce repeated computation and data movement while enabling persistent, long-running, and cross-model LLM applications.\color{black}

\subsection{LLM for Automatic Cloud Management}
\label{sec:llm-cloud-management}

\color{blue}Whereas Section 6.3 focuses on embedding LLM reasoning into individual system-software functions such as scheduling and resource allocation, this direction considers the broader question of whether LLM-enabled agents can participate in closed-loop management of complete cloud infrastructures.\color{black}

LLMs could be integrated into cloud management systems to automatically optimize the allocation of resources, monitor workloads, and dynamically adjust compute resources based on the system's needs. This would enable the cloud infrastructure to run autonomously, with LLMs making real-time decisions on resource provisioning, scaling, and maintenance, improving both performance and cost efficiency. Traditionally, cloud orchestration relies on static, rule-based autoscalers (such as the Kubernetes Horizontal Pod Autoscaler~\cite{kubernetes}) that reactively adjust to predefined metric thresholds like CPU or memory pressure. However, these heuristic methods often struggle to manage the bursty, multi-tenant, and highly variable nature of modern microservice workloads. By shifting toward an  AgentOps  paradigm, cloud-native infrastructures can leverage the semantic reasoning and predictive capabilities of LLMs to transition from reactive scaling to proactive, intelligent resource management.

Recent research highlights the effectiveness of multi-agent LLM frameworks in managing the complete lifecycle of cloud operations. For instance, the AIOPSLAB framework~\cite{chen2025aiopslab} demonstrates how specialized AI agents can autonomously execute complex operational tasks, from fault injection and anomaly detection to root-cause diagnosis and auto-remediation, directly within realistic cloud environments. Instead of treating monitoring data strictly as raw numbers, these agents apply semantic reasoning to unstructured operational telemetry (such as application logs, distributed traces, and configurations) to contextualize system health. Similarly, advanced frameworks like LADs~\cite{khan2025lads} utilize RAG and Chain-of-Thought reasoning to automate deployment pipelines; they generate accurate infrastructure settings and iteratively learn from deployment failures to continuously refine cloud management policies.

To ensure these autonomous decisions meet strict SLOs without excessive resource over-provisioning, state-of-the-art orchestration systems combine LLM capabilities with rigorous mathematical evaluation and strategic search models. For example, recent developments have integrated Monte Carlo Tree Search (MCTS) into multi-agent AIOps systems, providing a structured fault-reasoning tree that minimizes LLM hallucinations while accurately localizing bottlenecks across complex microservice topologies~\cite{ren2025multi}. Furthermore, studies utilizing microservice simulation environments validate that integrating LLM-driven reinforcement learning policies allows cloud managers to successfully solve constrained optimization problems, balancing cost-efficiency, low latency, and reliability, far more effectively than legacy heuristics~\cite{microservices}. By embedding these intelligent controllers directly into the underlying cloud substrate, infrastructure systems can autonomously self-heal and scale, embodying the ultimate vision of self-creating orchestration.

For LLM-driven cloud management to achieve widespread enterprise adoption, future research must urgently address the verifiable safety, explainability, and auditability of autonomous orchestration actions. A critical open problem is designing deterministic software guardrails that strictly bound the blast radius of AI agents, ensuring that an LLM's hallucination or miscalculation cannot trigger cascading infrastructural downtime. To accomplish this, the distributed systems community must develop standardized, high-fidelity cloud digital twins and continuous integration simulation environments. In these environments, LLM-generated management policies can be safely stress-tested against extreme edge-case failure injections, formally verified for SLO compliance, and mathematically audited before being granted execution privileges in live, multi-tenant production clusters.

\subsection{Decentralized LLM Infrastructure and Intelligence}
\label{sec:decentralized-intelligence}

\color{blue}Decentralized intelligence refers to distributing the intelligence and decision-making processes across multiple devices or nodes, rather than relying on a centralized system. The central research question is not simply whether intelligence can be decentralized, but how cloud-native LLM infrastructure should evolve when computation, model state, data, and control are distributed across heterogeneous administrative and geographic domains. This direction is motivated by the increasing cost and concentration of centralized LLM infrastructure, together with the privacy, resilience, and edge-computing requirements identified in Sections~3--5.\color{black}

Recent systems research has demonstrated that with intelligent orchestration, these distributed environments can rival centralized clouds in efficiency. For example, the Parallax system~\cite{tong2025parallax} introduces a two-phase decentralized scheduler that efficiently places LLM layer replicas across diverse volunteer nodes, optimizing both latency and throughput under strict memory and link-bandwidth constraints. To further address the network latency that often dominates these distributed setups, frameworks like Decentralized Speculative Decoding (DSD) turn communication delay into useful computation by verifying multiple candidate tokens in parallel across distributed nodes, yielding over 2.5$\times$ speedups in decentralized inference without requiring model retraining~\cite{song2025speculative}. Supporting these massive models on local edge devices also requires aggressive hardware-software co-design; innovations such as Resistive Random-Access Memory (RRAM)-based Processing-In-Memory (PIM) accelerators provide the massive internal bandwidth necessary to support low-latency edge LLM inference~\cite{wang2025low}. Furthermore, system models like the semi-Markov energy harvesting framework ensure that collaborative inference across battery-powered edge networks remains sustainable despite strict energy limitations~\cite{khoshsirat2024decentralized}.

Beyond pure inference routing, decentralization is revolutionizing how LLMs collaborate and train. Advanced multi-agent frameworks like AgentNet eliminate the traditional centralized orchestrator, introducing a decentralized, RAG-based architecture where agents autonomously adjust their connectivity and route tasks within a dynamically structured network. Similarly, Multi-Agent Actor-Critic (MAAC) methodologies enable independent LLM agents to execute long-horizon tasks based purely on local observations, avoiding the privacy risks of centralized execution~\cite{liu2026learning}. On the training side, Federated Large Language Models (FedLLMs) allow clients to collaboratively fine-tune models using low-rank adapters (LoRA)~\cite{lora}. To secure these distributed pipelines against malicious clients, novel defense mechanisms like Safe-FedLLM perform probe-based discrimination to filter out adversarial updates, ensuring decentralized intelligence remains both private and robust~\cite{tao2026safe}.

A critical future research direction for decentralized LLM orchestration is the development of robust, cost-aware verification mechanisms for permissionless environments. In a fully decentralized inference network where heterogeneous nodes (such as consumer GPUs or edge devices) process tokens, verifying the mathematical accuracy and quality of the generated outputs without re-executing the entire computation is a massive challenge. Future systems must formalize lightweight  Proof of Quality  (PoQ) consensus protocols that dynamically reward untrusted nodes based on a delicate balance of execution latency, energy efficiency, and semantic fidelity (e.g., using lightweight learned evaluators)~\cite{tian2025design}. Additionally, addressing the severe network bottlenecks introduced by cross-node tensor parallelism will require novel, asynchronous communication primitives designed specifically to tolerate the high latency and packet loss inherent in global, decentralized networks.

\subsection{Incentive Models for LLM Training}
\label{sec:incentive-models}

\color{blue}Within decentralized LLM infrastructures, incentives are not only an economic concern but also a systems mechanism for attracting, allocating, and regulating distributed data and computing resources. When infrastructure spans independently operated nodes, cloud providers, edge devices, and data owners, resource availability and system reliability depend on whether participants are appropriately rewarded for useful contributions and penalized for malicious or low-quality behavior.\color{black}

Developing fair incentive models begins with accurately quantifying the utility of individual contributions within massive, heterogeneous datasets. In the context of LLM instruction tuning, data valuation has emerged as a rigorous economic mechanism to determine fair compensation. For instance, the LimaCost framework~\cite{moon2025limacost} actively quantifies the quality of each data point by measuring its direct influence on model parameter updates during the alignment process, proving that incentivizing a small fraction of high-value data yields better instruction-following capabilities than training on massive, unfiltered datasets. To scale this concept across multiple untrusted data providers, systems are adapting cooperative game theory concepts like the Shapley value. Because exact Shapley calculation is prohibitively expensive, the NESTLE framework introduces a robust approximation using gradient tracing, allowing decentralized platforms to efficiently rank data contributions across diverse domains. Building on these valuations, recent works propose Fairshare pricing models~\cite{zhang2025fairshare} that align market incentives between LLM builders and human annotators. Empirical studies demonstrate that equitable, valuation-based compensation prevents the exploitation of data contributors, guaranteeing a sustainable supply of high-quality training data and preventing long-term market collapse.

Beyond data provision, robust incentive structures are essential to secure decentralized, federated, and multi-agent computational resources. When training models across permissionless networks, systems must reliably reward genuine computational effort while penalizing adversarial updates. The Gauntlet protocol~\cite{lidin2025incentivizing} exemplifies this by deploying an incentive system on a blockchain to evaluate pseudo-gradient contributions from anonymous peers, securely rewarding them based on their direct impact on minimizing the model's loss. Furthermore, mitigating self-interested behaviors, such as free-riding and knowledge hoarding, is critical in Cross-Silo Federated Learning. Recent frameworks introduce dynamic participant screening mechanisms and confidence attenuation monitoring to isolate and penalize clients who attempt to benefit from the global foundation model without contributing high-quality local updates~\cite{zhang2025incentive}. Similarly, in privacy-sensitive federated environments, the WinFLoRA framework~\cite{kou2026winflora} introduces a noise-aware incentive mechanism for Federated LoRA; it up-weights low-noise local updates, aligning individual participant utility with the global model's performance objective while accommodating heterogeneous privacy constraints.

A critical future research direction in LLM incentive modeling is the development of real-time, Byzantine-fault-tolerant attribution mechanisms that operate seamlessly across multi-modal streaming data. As AI systems increasingly train on continuous, live data (e.g., decentralized video feeds, edge sensor arrays, and real-time human feedback), batch-processed Shapley valuations will become obsolete. Future cloud-native orchestrators must integrate streaming data valuation algorithms natively into their ingestion pipelines, calculating cryptographically verifiable micro-rewards on a per-token or per-frame basis. Furthermore, as global AI regulations strictly enforce the right to be forgotten, incentive structures must be mathematically entangled with machine unlearning capabilities. Researchers must design dynamic smart contracts that automatically claw back or recalculate compensation when a data contributor invokes their right to withdraw data, ensuring that the economic ledger accurately reflects the model's current, legally compliant knowledge base.

\subsection{Selective Semantic Extraction}
\label{sec:semantic-extraction}

\color{blue}Selective semantic extraction becomes an infrastructure problem when cloud-native LLM services need to decide which model and execution path should process each part of a complex information-processing workload. Relying on a large foundation model for every extraction operation can create unnecessary accelerator utilization, latency, energy consumption, and cost. The infrastructure-level opportunity is therefore to decompose semantic workloads and dynamically route them among heterogeneous LLMs, SLMs, retrieval components, and deterministic processing services.\color{black}

Moving away from unstructured text generation toward rigid, auditable data extraction has led to the development of systematic frameworks that enforce strict structural schemas. As LLMs are increasingly deployed to automate enterprise-scale workflows, extracting structured information (e.g., deeply nested JSONs) from unstructured documents has become a mission-critical capability. However, recent benchmarking efforts, such as the ExtractBench~\cite{ferguson2026extractbench}, reveal that even frontier models suffer sharp degradation when required to execute large-scale schema extraction from complex PDFs. This highlights the urgent need for evaluation methodologies that score semantic equivalence and array matching rather than just valid syntax. To address these reliability gaps, systems like PARSE~\cite{shrimal2025parse} optimize JSON schemas directly for LLM consumption, creating a virtuous cycle where clear structural contracts drastically reduce extraction hallucinations and schema non-adherence. Furthermore, systematic toolkits like DELM (Data Extraction with Language Models)~\cite{fithian2025delm} and benchmarking suites like LLMStructBench~\cite{tenckhoff2026llmstructbench} formalize extraction as a predictive modeling task, utilizing constrained decoding and token masking to guarantee valid structured outputs directly from the model's output stream.

Beyond schema compliance, achieving high precision in deep knowledge discovery tasks requires embedding structured constraints directly into the model's decoding and reasoning mechanisms. For example, recent research introduces Structure-Aware Decoding Mechanisms~\cite{qiu2025structure}, which solve the complex challenge of overlapping and nested entity extraction by integrating hierarchical structural constraints during the decoding phase. By capturing multi-granular entity span features through structured attention modeling, the system dynamically enforces consistency between the extracted semantics and the required output format. Additionally, multi-agent pipelines are being leveraged to guarantee semantic fidelity; frameworks like Manalyzer~\cite{xu2025manalyzer} employ hierarchical extraction paired with automated self-proving and feedback mechanisms to actively mitigate hallucinations during the extraction of complex tabular and numerical data from scientific literature. Similarly, the Hierarchical Semantic Piece (HSP) framework~\cite{liu2025reducing} extracts multi-granularity semantic elements, using sentence-level pieces for global context and entity-level pieces for local details, allowing orchestration systems to autonomously cross-reference and filter out factual inconsistencies.

A critical future research direction in selective semantic extraction is the development of ultra-efficient, heterogeneous Small Language Model (SLM) routing architectures specifically optimized for deterministic token extraction. Relying entirely on massive foundation models for token-level entity matching is often computationally prohibitive, cost-inefficient, and prone to severe latency bottlenecks. Future cloud-native orchestration systems must deploy hybrid, agentic architectures where a large, generalist LLM acts strictly as an orchestrator, interpreting user intent and defining the extraction schema, while delegating the actual selective semantic extraction to a fleet of specialized, CPU-efficient SLMs. Systems research must address how to seamlessly execute this macro-to-micro handoff, ensuring that exact character offsets, mathematical confidence scores, and strict structural formats are securely preserved and aggregated across distributed microservices without inducing cross-model context dilution or data leakage.

\subsection{LLM@Home: Edge-Cloud and Multi-Device Infrastructure}
\label{sec:llm-home}

\color{blue}From the perspective of this research agenda, LLM@Home should not be interpreted merely as running a smaller model on a personal device. Instead, it represents an extreme case of distributed edge-cloud infrastructure in which computation, memory, contextual state, and control must be coordinated across highly heterogeneous and intermittently available devices under strict privacy and energy constraints. The fundamental research problem is therefore how to construct a secure and adaptive local AI infrastructure from a dynamic collection of resource-constrained devices.\color{black}

As smart home ecosystems and the Internet of Things expand, relying on centralized cloud servers for every interaction introduces unacceptable latency, creates single points of failure, and poses severe privacy risks. By moving inference to the network edge, the LLM@Home paradigm ensures that sensitive personal data, such as voice recordings, daily routines, and home security telemetry, never leaves the local environment. Recent applied research demonstrates that on-device LLMs can effectively unify natural language command interpretation and home automation execution without relying on proprietary cloud APIs. For instance, a recent study on fine-tuning 0.5B-parameter Small Language Models (SLMs) for home assistants proves that an entirely local, CPU-based model can accurately perform complex slot and intent detection alongside natural response generation, matching the utility of cloud solutions while ensuring total user privacy~\cite{birkmose2025device}. Furthermore, innovative frameworks like Vega~\cite{al2025vega} and VoiceTalk~\cite{lin2025voicetalk} have successfully integrated LLMs with smart home microservices, allowing users to dynamically control heterogeneous IoT devices via intuitive natural language without writing complex automation code.

However, deploying these computationally demanding models onto resource-constrained edge hardware, such as Raspberry Pis, smart speakers, or mobile phones, presents extreme technical hurdles. To bridge the gap between model size and available memory, aggressive system-level optimizations are essential. Cutting-edge model compression techniques, including 4-bit and 8-bit quantization and parameter pruning, have made it possible to fit highly capable models into limited RAM without catastrophic performance degradation. For example, recent empirical evaluations using the LEAF (LLM Edge Assessment Framework) demonstrate that applying localized quantization and memory-aware offloading allows even repurposed, low-power edge GPUs to achieve highly efficient token-per-second generation rates for smart home reasoning tasks~\cite{abdulkadhim2026introducing}. Additionally, emerging algorithmic accelerations, such as staged speculative decoding and token tree collaboration, allow edge devices to rapidly verify tokens in parallel, accelerating real-time inference latency by up to 9x compared to standard decoding methods~\cite{xu2024device}.

A critical future research direction for the LLM@Home paradigm is enabling seamless, secure multi-device collaboration within transient edge networks. As users accumulate heterogeneous AI-enabled devices (e.g., smartphones, smart TVs, autonomous delivery robots, and wearables), relying on isolated on-device models will limit contextual awareness. Future orchestration frameworks must develop decentralized  Edge-Mesh  protocols that allow local LLMs to dynamically share context, negotiate access control policies, and pool computational resources across the home network without centralized coordination. For instance, transient device collaboration systems point toward a future where LLMs autonomously generate real-time, fine-grained access control policies as new, unfamiliar devices enter the home environment~\cite{shastri2025llm}. Systems research must formalize these distributed, peer-to-peer LLM communication protocols while guaranteeing mathematical bounds on energy consumption and data privacy.

\subsection{Personalized LLM}
\label{sec:personalized-llm}

\color{blue}Personalization changes the unit of resource management from a shared model to a shared model plus per-user state\color{black}. Personalized LLMs would involve adapting the model to an individual user's preferences, knowledge, and interactions. These models would learn from user behavior over time, allowing them to provide more relevant and accurate responses. Personalization could be applied to various domains, such as virtual assistants, recommendation systems, and content creation tools.

From a systems and orchestration perspective, serving personalized models at a massive scale introduces severe memory and scheduling bottlenecks. If every individual user requires a distinctly fine-tuned model, hosting them in the cloud quickly violates infrastructure memory limits. To resolve this, modern cloud-native systems rely on Parameter-Efficient Fine-Tuning (PEFT) and multi-tenant serving architectures, where lightweight adapters (e.g., LoRA~\cite{lora}) are dynamically swapped on top of a frozen, shared base model. However, even the  One-PEFT-Per-User  paradigm requires computationally intensive backend training. To overcome this, cutting-edge architectures like Profile-to-PEFT (P2P)~\cite{tan2025instant} introduce scalable hypernetworks trained end-to-end to map a user's encoded profile directly into a full set of personalized adapter parameters in a single forward pass. This entirely eliminates the need for per-user training at deployment, enabling instant, real-time LLM personalization at an industrial scale while drastically reducing computational overhead. Similarly, the Embedding-to-Prefix framework~\cite{huber2025embedding} demonstrates how systems can dynamically generate user-specific prompt prefixes from latent memory embeddings without modifying the backbone LLM, achieving deep personalization with minimal memory footprint.

At the infrastructure level, hosting multi-tenant personalized applications, where different users generate vastly different sequence lengths and request volumes, creates complex fairness and resource starvation challenges. The FAIRSERVE system~\cite{khan2024ensuring}, developed following extensive analysis of platforms like MS Copilot, introduces Overload and Interaction-driven Throttling (OIT) specifically designed for multi-tenant LLM serving. By tracking a Weighted Service Counter across distributed GPU clusters, it ensures fair access and prevents token wastage without degrading the latency of highly personalized user interactions. Furthermore, personalization is increasingly shifting toward dynamic, real-time adaptation. Test-time personalization algorithms like T-POP~\cite{qu2025t} synergistically combine online preference feedback with dueling bandits, enabling the LLM to personalize its decoding strategy and align with user preferences after just a few rapid interactions, bypassing the need for heavy, batch-processed retraining pipelines altogether.

\color{blue}From an infrastructure perspective, personalization changes the unit of resource management from a model instance to a combination of shared model parameters and dynamically evolving user-specific state. Future research must therefore investigate how millions of adapters, prefixes, memories, and personalized KV states can be efficiently stored, cached, loaded, scheduled, updated, and removed. Important questions include how personalized workloads should share accelerators without creating unfairness or interference, how strong isolation can be maintained without sacrificing batching efficiency, how user state should migrate across distributed replicas, and how machine unlearning can remove personalized information from shared infrastructure. Addressing these issues would enable large-scale personalization while preserving the efficiency, fairness, and privacy requirements of multi-tenant cloud services.\color{black}

\subsection{Future Heterogeneous Computing Paradigms (Quantum/Neuromorphic)}
\label{sec:quantum-computing}

\color{blue}
The relevance of quantum computing to the present research agenda therefore lies less in treating quantum processors as near-term replacements for GPUs and more in understanding how future cloud-native infrastructures could incorporate fundamentally heterogeneous computing substrates. The corresponding systems challenge is to determine how classical accelerators and emerging quantum resources can be jointly exposed, scheduled, and coordinated within distributed runtimes.

Representative open research questions include which LLM subroutines are sufficiently self-contained to justify offloading to quantum resources; how execution should be partitioned between classical and quantum processors under communication and synchronization overheads~\cite{liang2023unleashing}~\cite{wan2023efficient}; how cloud-native resource managers should describe the capabilities, availability, and uncertainty of quantum resources; and how hybrid workflows should be evaluated against classical baselines in terms of end-to-end latency, energy consumption, reliability, and cost. Because current quantum hardware remains limited, these questions represent a longer-term research agenda. Nevertheless, the underlying problems of heterogeneous resource abstraction, workload decomposition, and cross-platform orchestration are directly relevant to the evolution of future distributed LLM infrastructures.\\

\textbf{Takeaway.} Taken together, the research directions in this section indicate that the evolution of LLM systems will require corresponding changes throughout the cloud-native and distributed infrastructure stack. The agenda extends from low-level computation and heterogeneous hardware, through system software and autonomous orchestration, to context and data management, decentralized execution, edge-cloud collaboration, and personalized multi-tenant services. Although these directions operate at different layers and exhibit different levels of technological maturity, they share a common systems requirement: future infrastructure must become increasingly aware of the computational structure, contextual state, service objectives, and deployment environment of LLM workloads.

The research agenda also suggests a progression in maturity. In the near term, advances in sparse computation, heterogeneous scheduling, prompt and context optimization, multi-tenant serving, and systematic infrastructure measurement can build directly on existing cloud-native platforms. In the medium term, distributed context management, LLM-assisted system software, autonomous cloud orchestration, edge-cloud collaboration, and decentralized resource coordination require deeper integration across system layers. Longer-term directions, including hybrid quantum-classical execution and highly decentralized AI infrastructures, will require new resource abstractions, programming models, and orchestration mechanisms. Across these horizons, the central objective remains consistent: transforming current cloud platforms from general-purpose computing infrastructures that host LLMs into LLM-aware infrastructures specifically designed to manage their computation, state, resources, and operational complexity.

\color{black}

\section{Summary and Key Findings}
\label{ch:7}

In this paper, we explored the critical role that cloud-native and distributed systems play in scaling LLMs. As the computational demands of LLMs continue to rise, traditional middleware and semantic models cannot adequately address the unique requirements of LLM applications. This necessitates the integration of advanced cloud-native architectures, distributed systems, and emerging technologies to manage and optimize these complex models.

\subsection{Recap of the Necessity for Cloud-Native and Distributed Systems in Scaling LLMs}
\label{sec:recap}

The rise of LLMs has introduced substantial computational intensity and operational complexity. Training processes depend on large-scale parallelism across accelerators and high-bandwidth interconnects, while inference services must handle bursty, user-driven workloads with strict latency requirements. Cloud-native systems provide elasticity, modularity, and automation, whereas distributed systems contribute scalable parallel execution, topology awareness, and fault tolerance.

The analysis indicates that neither paradigm alone is sufficient; rather, their integration forms the foundation for supporting LLM workloads at scale. LLM execution exhibits phase-level heterogeneity, such as distinctions between prefill and decode stages, and operator-level differences between attention and feed-forward computation. These characteristics necessitate fine-grained orchestration strategies that align workload structure with heterogeneous hardware capabilities.

Key system-level pressures include the need to optimize across networks, accelerators, memory hierarchies, storage systems, and microservice-based architectures. Variability in prompt length, decoding behavior, and multi-tenant demand further complicates scheduling and autoscaling decisions.

A principal conclusion is that resource management for LLMs must evolve from coarse-grained, utilization-driven policies toward structure-aware and topology-aware orchestration mechanisms. Hardware--LLM co-design, elastic scheduling across heterogeneous accelerators, and integration of cloud--edge continuum are not optional optimizations but foundational requirements for predictable performance and cost control. Innovations such as serverless inference and function-as-a-service paradigms further illustrate how cloud-native abstractions can be extended to accommodate stateful, accelerator-intensive AI services.

\subsection{Vision: Democratizing Access to Foundation Models Through Robust Infrastructure}
\label{sec:vision}

\color{blue}The research agenda developed in Section~6 further suggests that realizing this vision will require coordinated advances across heterogeneous computation, intelligent orchestration, distributed context and state management, decentralized edge-cloud execution, and personalized multi-tenant services.\color{black}

The long-term vision is to establish a robust, scalable, and sustainable infrastructure capable of democratizing access to foundation models. By lowering operational complexity and improving portability across environments, cloud-native and distributed systems can reduce barriers to entry for academic, industrial, and public-sector stakeholders.

Democratization in this context does not imply simplification of underlying complexity, but rather the development of abstractions that encapsulate sophisticated resource management, privacy enforcement, and sustainability considerations within programmable control planes. Such infrastructure must integrate elasticity, multi-tenant fairness, energy efficiency, and carbon awareness as first-class objectives alongside latency and throughput.

Emerging technologies such as serverless computing, federated and decentralized training, reinforcement learning--driven orchestration, and hybrid cloud--edge deployments suggest that future LLM platforms will be geographically distributed, heterogeneity-aware, and policy-driven.

Furthermore, the integration of quantum computing, edge computing, and reinforcement learning agents into LLM infrastructure can unlock new opportunities for optimization and innovation. These technologies will enable LLMs to operate more efficiently, adapting dynamically to changing workloads and enhancing the flexibility of AI systems at scale.

\subsection{Call for Academic, Industrial, and Policy Collaboration}
\label{sec:collaboration}

As the LLM ecosystem grows, it is essential for academic institutions, industry leaders, and policy makers to collaborate closely. Academic research can drive innovation in LLM algorithms and architectures, while industry can contribute to the real-world implementation and scaling of these systems. Policy makers must play a role in ensuring ethical standards and data privacy protections as LLMs become more prevalent in global applications.

There is a pressing need for standardization of LLM deployment techniques, training APIs, and resource management protocols to facilitate interoperability across diverse platforms and cloud providers. Collaboration across these sectors will help address the challenges faced by LLMs, particularly around autoscaling, data management, and resource allocation, ensuring that LLMs continue to evolve into a sustainable and impactful technology for the future.

\subsection{Key Contributions and Findings}
\label{sec:contributions}

This paper has outlined the current state of LLM technology, the challenges faced by cloud-native and distributed systems, and the potential solutions to these challenges. Key contributions include:

\begin{itemize}
    \item An analysis of the computational, system, and operational challenges in supporting LLMs at scale.
    \item An exploration of cloud-native and distributed architectures that enable LLM scaling, including microservices, containerization, and orchestration.
    \item A discussion of emerging trends and innovations such as quantum computing, federated learning, and serverless inference that could shape the future of LLMs.
    \item A research agenda for the development of LLM infrastructure, with recommendations for collaboration between academia, industry, and policy makers.
\end{itemize}

\section{Conclusions}
\label{ch:8}

In summary, the continued scaling and optimization of large language models will fundamentally depend on advances in cloud-native architectures, distributed systems, and emerging computing paradigms. As LLM workloads grow in complexity and scale, traditional infrastructure alone is unlikely to sustain the required levels of efficiency, adaptability, and performance. Instead, tightly integrated system innovations, spanning hardware, runtime, and orchestration layers, will be essential to fully realize the potential of next-generation AI systems. With a robust and intelligently designed infrastructure, LLMs are poised to unlock transformative capabilities across a wide range of domains and industries.

Achieving this vision, however, requires sustained progress across several critical research agendas. First, improving computational efficiency remains a central challenge, calling for advances in training and inference algorithms, deeper hardware–software co-design, and the exploration of novel acceleration techniques. Second, the development of system software must evolve toward LLM-aware abstractions and cross-layer optimization frameworks, supported by enhanced observability and performance diagnostics to manage increasingly complex workloads.

In addition, resource management strategies must become significantly more adaptive and fine-grained, incorporating predictive and multi-objective scheduling mechanisms that jointly optimize for performance, energy consumption, and operational cost. At the same time, privacy and security considerations will play an increasingly important role, necessitating robust privacy-preserving learning techniques, secure execution environments, and proactive defenses against emerging system-level and model-level threats. Furthermore, the establishment of widely accepted standards, including unified interfaces, benchmarking methodologies, and best practices, will be critical to ensuring interoperability, reproducibility, and fair evaluation across platforms and deployments.

Ultimately, addressing these challenges will require close collaboration across academia, industry, and government. By fostering such coordinated efforts, it will be possible to build a sustainable, secure, and equitable ecosystem for LLMs, one that not only advances the state of AI, but also delivers broad societal benefits in a responsible and scalable manner.

\section*{Acknowledgement}
This work is supported by the National Key R\&D Program of China (No. 2026YFE0199800), National Natural Science Foundation of China under Grant 62572462, Guangdong Science and Technology Cooperation Project (No. 2025A0505020065), Key Research and Development and Technology Transfer Program of Inner Mongolia Autonomous Region (2025YFHH0110) and Shenzhen Science and Technology Program under Grants JCYJ20240813155810014 and ZDYJ20251211121533004.

\bibliography{ref}

\end{document}